\documentclass[reprint,
 amsmath,amssymb,
 aps,
pra,
]{revtex4-2}

\usepackage{graphicx}
\usepackage{dcolumn}
\usepackage{bm}

\begin{document}

\preprint{APS/123-QED}

\title{Activity-Induced Stiffness, Entanglement Network and Dynamic Slowdown in Unentangled Semidilute Polymer Solutions}

\author{Jing Li}
 \affiliation{School of Physical Science and Technology, Southwest University, Chongqing 400715, China}
\affiliation{Chongqing Key Laboratory of Micro-Nano Structure Optoelectronics, Chongqing 400715, China}
\author{Bokai Zhang}
\email{zbk329@swu.edu.cn}
 \affiliation{School of Physical Science and Technology, Southwest University, Chongqing 400715, China}
\affiliation{Chongqing Key Laboratory of Micro-Nano Structure Optoelectronics, Chongqing 400715, China}
\author{Zhi-Yong Wang}
\email{zywang@swu.edu.cn}
 \affiliation{School of Physical Science and Technology, Southwest University, Chongqing 400715, China}
\affiliation{Chongqing Key Laboratory of Micro-Nano Structure Optoelectronics, Chongqing 400715, China}
\date{\today}

\begin{abstract}
Active polymers possess numerous unique properties that are quite different from those observed in the system of small active molecule due to the intricate interplay between their activity and topological constraints.  This study focuses on the conformational changes induced by activity, impacting effective stiffness and crucially influencing entanglement and dynamics.   When the two terminals of a linear chain undergo active modification through coupling to a high-temperature thermal bath,  there is a substantial increase in chain size, indicating a notable enhancement in effective stiffness.  
Unlike in passive semiflexible chains where stiffness predominantly affects local bond angles, activity-induced stiffness manifests at the scale of tens of monomers.  While activity raises the ambient temperature, it significantly decreases diffusion by over an order of magnitude. The slowdown of dynamics observed can be attributed to increased entanglement due to chain elongation. 
\end{abstract}

\maketitle

\section{Introduction}
Active polymers are characterized by the presence of specific segments that possess the ability to harness energy from their surrounding environment at a particle level, thereby converting it into their own kinetic energy.   The influx and dissipation of energy on a microscopic scale are closely associated with  fundamental biological processes, such as DNA transcription\cite{Goloborodko2016},  colocalization and segregation of active DNA \cite{Jingyu2019,Zhiming2011,Bohrer2023,Stephanie2013,Ganai2014,Nuebler2018,Mahajan2022},  micron-scale organization of interphase chromatin\cite{Saintillan2018, Zidovska2013}, and activity-induced genome conformations\cite{Goychuk2023,Misteli2020}.  In addition , the investigation of active polymers has stimulated the synthesis of active artificial materials\cite{Biswas2017,Leonardo2016,Xinlong2022,Yan2016}, revealing significant potential for applications in microrobots, micromixing, and targeted drug delivery\cite{Stefano2018,Park2017,Ober2015}. 

The nonequilibrium systems of active polymers exhibit a myriad of phenomena distinct from those observed in equilibrium systems and elude prediction and explanation by conventional polymer physics theories.  For example,  in equilibrium, achieving conventional glass transition in a melt of ring polymers necessitates a higher concentration or lower temperature. \cite{Gao2019,George2018}  However, in out-of-equilibrium active rings, the introduction of active segments within a ring induces long-lived interlockings among different rings due to topological constraint, leading to a dynamically arrested state labelled as active topological glass \cite{Smrek2020,Ubertini2022,Micheletti2024}.  
Similar to the phase separation of polymer mixtures at equilibrium, the presence of activity difference in a mixture of active and passive polymers is considered to mimic the Flory interaction parameter and induce polymer segregation\cite{Smrek2017PRL}.  This provides a potential explanation for the three-dimensional spatiotemporal organization of the genome inside the cell nucleus, where non-equilibrium ATP-powered processes govern chromatin dynamics\cite{Saintillan2018,Bruinsma2014}. The interplay between activity and chain connectivity gives rise to intriguing and distinctive movements in various environments, such as railway\cite{Li2023}, helical\cite{Anand2018}, and snake-like motion\cite{Rolf2015,Rolf2016}, contrasting with the behavior of small active particles.  Active monomers have been observed to induce globule-to-coil or coil-to-globule transitions depending on their self-propelled force \cite{Jain2022,Natali2020,Bianco2018,Yang2023,Yan2023}.  When the motor of activity is introduced to the central monomer, the radius of gyration decreases, accompanied by a significant increase in chain diffusion by several orders of magnitude\cite{Davide2011,Davide201102}. Polymers with active terminal leads to enhanced persistence lengths and increased effective stiffness, as previously observed in simulation studies\cite{Li2023,Natali2020}.  { The influence of the position of active segments along a chain contour on isolated partially active polymer chains with tangential self-propelled force was investigated by Vatin \textit{et al}. \cite{Locatelli2024}  It was found that when the active segments are located at one of the ends of the polymer, the chain size increases and its dynamics slow down.   The presence of an active force acting tangentially to the polymer contour in entangled polymer melts is observed to result in a significantly higher shear stress plateau, accompanied by a reduction in the end-to-end distance. \cite{Davide2023}  The entanglement of passive polymer chains in isotropic melts increases with stiffness,\cite{Moreira2015,Svaneborg2020} and as the stiffness further increases, at the onset of the isotropic-nematic transition, the entanglement length reaches its minimum. \cite{Hoy2020}  In semidilute polymer solutions, molecular dynamics simulations and comprehensive scaling theory also demonstrated that entanglement increases proportionally to the Kuhn length.  \cite{Milner2020,Everaers2008}

In cellular chromosomes, there are typically active and inactive monomers that organize into euchromatin and heterochromatin compartments based on gene density. \cite{Mahajan2022}  These ATP-dependent enzymatic activities associated with local chromatin remodeling and transcription give rise to stochastic forces.  The nonequilibrium stochastic force fluctuations can be achieved by a local effective temperature, where active monomers can be regarded as ‘hot’ particles while inactive particles can be considered ‘cold’ particles. \cite{Ganai2014} Therefore, a natural and yet unexplored question lies in the effect of active modification and its distributions on chain size, entanglement network and dynamics in unentangled semidilute solution, which is closely related to chromosome conformation, organization and dynamics. \cite{Pilar2020,Clapier2017,Thirumalai2021}
}

Here, we focused on the conformation, entanglement, and dynamic properties influenced by active monomers in unentangled semidilute polymer solutions.  Our aim is to address these fundamental questions: How does the incorporation of high-temperature active monomers impact the conformations of polymer chains? What is the influence of the positioning of active monomers on the extent of chain expansion? Can active modification on polymer chains induce effective entanglement in unentangled polymer solutions? Does activity consistently enhance mobility in such a system?

\section{Model and Simulation Method}\label{sec:2}
\begin{figure*}[!t]
\centering
	\includegraphics[width=0.8\linewidth]{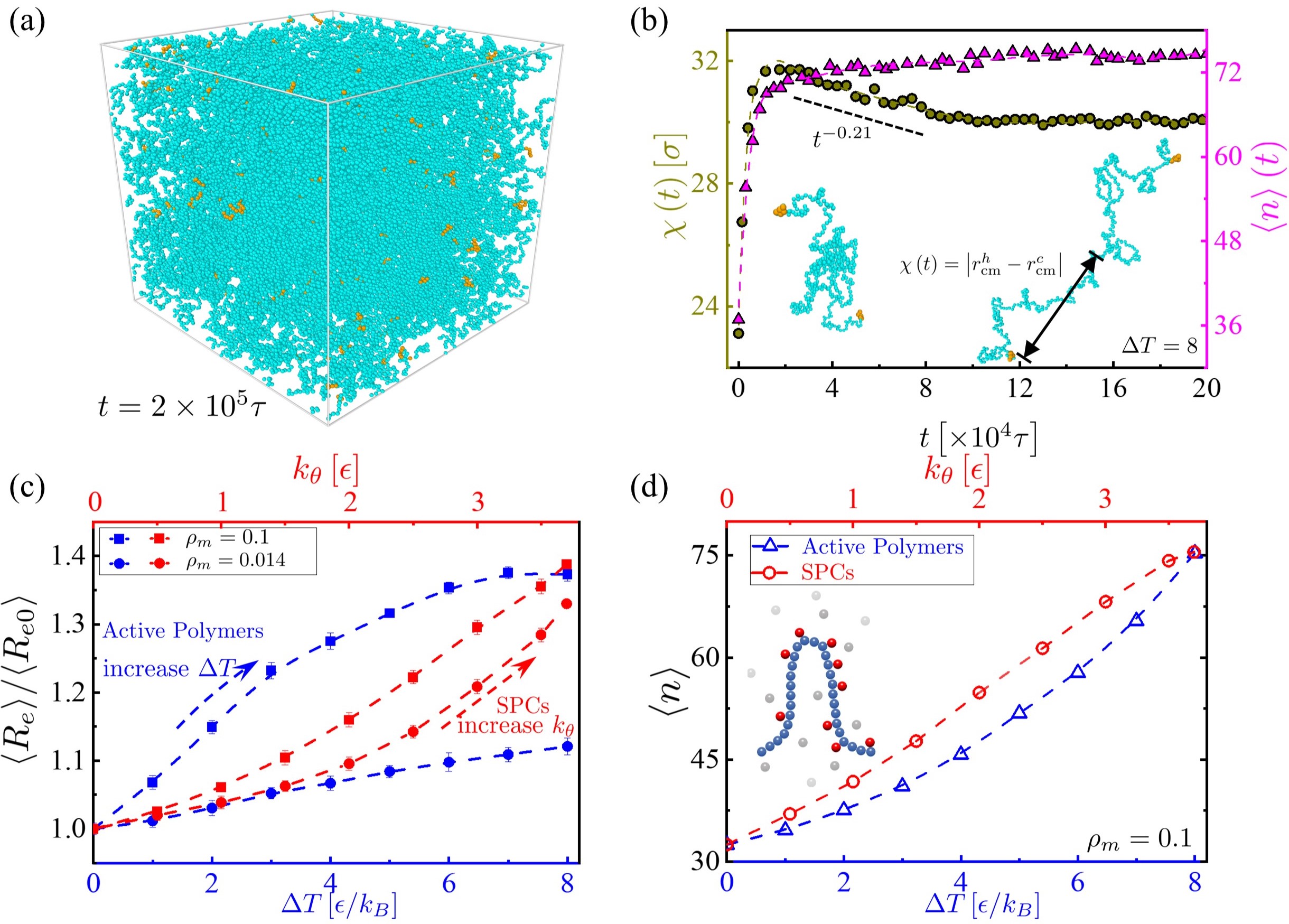}
    \caption{(a) Snapshots of the system at $t=2\times 10^5\tau$ after the onset of activity. (b) Time evolution of deviation function (dark yellow) an contact number (magenta) at activity difference $\Delta T=T_h-T_c$.  Insets: typical snapshots of active polymers in equilibrium ($t=0$) and steady state ($t=2\times10^5\tau$).  The active monomers on the terminals are shown in orange and the passive monomers on the center are shown in blue. The dashed lines are guided to the eyes.  (c) The end-to-end distance for the system of active polymers at various activity differences (the blue $x$-axis on the bottom) and for the system of SPCs at various bending energy coefficients (the red $x$-axis on the top).  Two monomer number densities are considered, semidilute solution with $\rho_m = 0.1$ (diamonds), and dilute solution with $\rho_m=0.014$ (circles).  (d) The average number of  contact monomers versus activity differences (the blue $x$-axis on the bottom) and versus bending energy coefficients (the red $x$-axis on the top) at monomer density $\rho_m=0.1$.  Inset: schematic of contact monomers around a tagged chain.}
    \label{fig:1}
\end{figure*}
To elucidate these questions, we conducted molecular dynamics simulations on semidilute polymer solution of fully flexbile Kremer-Grest bead-spring model\cite{Kremer1990}, of $M=100$ chains of $N=500$ beads with diameter $\sigma$.   All monomers interact via a purely repulsive shifted Lennerd-Jones potential (WCA potential)
\begin{equation}
U_{\mathrm{LJ}}(r)=\Bigg(4\epsilon \Bigg[ \Big(\frac{\sigma}{r}\Big)^{12}-\Big(\frac{\sigma}{r}\Big)^6 \Bigg]+\epsilon\Bigg)\Theta(2^{1/6}\sigma-r)
\label{eq:1}
\end{equation}
where the Heaviside step function $\Theta$ ensures that the potential is cut at $2^{1/6}\sigma$.  The chain connectivity was achieved through a combination of a repulsive WCA potential and the finitely extensible nonlinear elastic (FENE) potential
\begin{equation}
    U_{\mathrm{FENE}} = -\frac{1}{2}KR^2_0ln\Big[1-\Big(\frac{r}{R_0}\Big)^2\Big]
\end{equation}
where $K=30\epsilon/\sigma^2$ and $R_0=1.5\sigma$.  { With the choice of these parameters, the combination of LJ potential and FENE potential has a minimum at $l_b=0.96\sigma$, thereby prohibiting chain crossings.  The steepness of FENE ensures that, within the temperature range under investigation, the fluctuations in bond length are approximately 1\%}   All quantities are given in LJ reduced units throughout the article.  The units of temperature, length, time, and mass are respectively $\epsilon/k_B$, $\sigma$, $\tau$, and $m$, where $\epsilon$ and $\sigma$ are defined by LJ potential (eq \ref{eq:1}), $k_B$ is Boltzman constant, $m$ is the mass of a monomer, and $\tau=\sqrt{m\sigma^2/\epsilon}$.  

The simulation was carried out in a periodic cubic box with a fixed monomer density.  { In this study, we mainly focus on semidilute solution with a number density of monomers $\rho_m=0.1$, which is greater than the overlap concentration estimated from the radius of gyration in an infinitely dilute solution.  In addition, we also consider dilute solution with a density of $\rho_m=0.014$, which falls below the overlap concentration and can be regarded as an isolated polymer chain. 
} For equilibrium polymers, the entanglement length in this polymer model is found to be $N_e\approx 85$ at $\rho_m=0.85\sigma^{-3}$, and $N_e$ scales approximately as $\sim \rho_m^{-2}$ \cite{Fetters1994, Martin2000, Milner1999}.  Therefore, these systems studied were always in the unentangled regime.  The monomer density $\rho_m=0.014\sigma^{-3}$ is lower than the estimated overlap volume fraction based on the polymer's radius of gyration at infinite dilution.  

We perform the simulations at constant volume and temperature with Langevin thermostats,
\begin{equation}
    m\ddot{\bm{r}}_i=-\nabla_iU(\{\bm{r}_i\})-m\Gamma \dot{\bm{r}}_i+\sqrt{2m\Gamma k_BT}\bm{\zeta}(t)
    \label{eq:3}
\end{equation}
where $\bm{r}_i$ is the position vector of $i$-th monomer, $U(\{\bm{r}_i\})$ is the total interaction potential acting on the monomer.  $\Gamma$ represents friction coefficient.  The last term of eq \ref{eq:3} is a random collision force depending on the temperature of solvent.  

Following Smrek \textit{et al.}'s approach\cite{Smrek2017PRL}, the monomers were considered as active or ‘hot’ particles by subjecting them to a Langevin thermostat with a higher temperature $T_h$, while the passive monomers were coupled to a second Langevin thermostat with a lower temperature, $T_c=1.0\epsilon/k_B$.  {In this study,  the distribution of active monomers along chains was extensively investigated.  We primarily reported the case where a consecutive sequence of 5 monomers at each end of the linear chain was selected as high-temperature active monomers, while considering the rest $N-10$ monomers in the center of the chain as passive ‘cold’ monomers with $T_c$.  The active modification at the chain ends has significant implications for both conformation and dynamics, akin to the active head of filamentous bacteria or worms. \cite{Deblais2020}}  Their snapshots are shown in the inset of Figure \ref{fig:1}b.  The systems investigated ranged from $T_h=1.0\epsilon/k_B$ to $9.0\epsilon/k_B$. 
{ The friction coefficient quantifies the heat transfer between hot and cold monomers.  When it exceeds a certain threshold, nonequilibrium phase separation occurs.  It has been demonstrated that longer chains exhibit a higher propensity for phase separation. \cite{Smrek2017PRL}  Here, the friction coefficient in the Langevin dynamics simulation was set to $0.02/\tau$,  ensuring that nonequilibrium phase separation is prevented. 
The monomers in the system remain homogeneous even after the onset of an activity lasting $2\times 10^5 \tau$ as shown in Figure \ref{fig:1}a.  }

For comparison, we also investigated the equilibrium system of semiflexible passive chains (SPCs) in a linear fashion with local bending energy $U_{\mathrm{BEND}}=k_{\theta}(1-cos(\theta-\pi))$.  We conducted the simulations at a fixed temperature $T=1.0\epsilon/k_B$ with varying values of $k_{\theta}$.  The other simulation parameters and interactions were maintained consistent with those used in the active polymer system. 

All simulations were conducted using the LAMMPS software package. \cite{Plimpton1995}  To initialize the system, the polymers are randomly placed within the box.  The overlap between particles is removed by employing a soft potential.  Subsequently, an equilibrium run is performed under the NVT ensemble with a time step of $\delta t=0.001\tau$ at $T=1.0\epsilon/k_B$ without any activity for all monomers.  After an equilibration period, we activated the two terminals of each linear chain for the system of active polymers at higher temperature or introduced local bending energy for the system of SPCs at time $t=0$.  The length of equilibration in our simulation exceeded several times the Rouse relaxation times (see Supplementary Information (SI)).   All quantities were averaged over a minimum of 10 independent replicas.  

\section{Results and Discussion}
\subsection{Activity-induced chain expansion}
The deviation between the active segments and cold segments of a chain can be characterized by a time-evolution function\cite{Smrek2020}, $\chi(t)$, as illustrated in the inset of Figure \ref{fig:1}b.  It is defined as  $\chi(t)=|\bm{r}_{\mathrm{cm}}^h-\bm{r}_{\mathrm{cm}}^c|$, where $\bm{r}_{\mathrm{cm}}^h$ and  $\bm{r}_{\mathrm{cm}}^c$ are the centers of mass of active ends and midchain passive segments, respectively.  Figure \ref{fig:1}b shows that $\chi(t)$ rapidly increases to a peak at $t_{\mathrm{max}}\approx 1.5\times 10^4\tau$ and subsequently exhibits a gradual decline at $\Delta T=T_h-T_c=8\epsilon/k_B$, during which the corresponding end-to-end distance $R_e(t)$ also significantly increases over time, reaching its maximum value simultaneously (see Figure S1 in the SI).   { The increase in the $\chi$ function corresponds to the elongation of the polymer chain due to activity difference.  Hot particles could explore more configuration space.  At long time, the self-avoidance effect of the chain leads to a higher probability of outward movement along the backbone for hot monomers at the chain ends, resulting in elongation of the entire chain.}
The function $\chi$ shows a power-law decay, $~t^{-0.21}$ after surpassing this peak, indicating a slight compression of the chains after significant stretching occurs.  The decay is attributed to topological constraints imposed by neighboring chains during chain stretching.   No reduction of chain size is observed at low polymer density ($\rho_m=0.014\sigma^{-3}$) as shown in Figure S2 in the SI, providing support for this perspective.  

{ As the active chain extends, the neighboring particles around it may influence its conformation through random collisions with the tagged chain.  We define the contact monomers as those within a distance of less than $1.5\sigma$ from the nearest monomer on the tracer chain and belonging to other chains, as shown in the inset of Figure \ref{fig:1}d.
Accompanying the increase of $\chi$, the number of the contact monomers $\langle n\rangle$ 
exhibits a rapid increase after the onset of the activity.  }
It reaches its maximum at the same time $t_{max}$ as the $\chi$ function and subsequently approaching saturation.   This intriguing synchronicity with the $\chi$ function suggests that the contact number increases with chain size, and the monomer collisions within neighboring chains induce further elongation of the chain.

At long times, the deviation function $\chi(t)$ and the end-end distance $R_e(t)$ reach a plateau.  We computed their mean values after $t=10^5\tau$.  The mean end-to-end distance $\langle R_e\rangle$ and the number of contact monomers $\langle n\rangle$ confirm the significant role of interchain topological constraints.  When the activity difference increases from $\Delta T=0$ to $8\epsilon/k_B$ at $\rho_m=0.1\sigma^{-3}$, the mean end-to-end distance shows a 37\% increase compared to the passive system ($\Delta T=0$) in Figure \ref{fig:1}c.  However, in dilute solution with $\rho_m=0.014\sigma^{-3}$, $\langle R_e\rangle$  only shows a 11\% increase.  This suggests that not only activity differences but also interactions with surrounding chains contribute to a substantial enlargement in chain size.

Drawing inspiration from numerous studies, the incorporation of activity induces effective chain stiffness, thereby resulting in an elongation of the polymer chains. Additionally, we examined the chain size of SPCs for comparison, where the real stiffness among the local bonds contributs to their elongation.  Figure \ref{fig:1}c shows that the chain size increases with bending energy coefficient. { However, unlike the activity-induced increase, this elongation does not reach saturation within the range of bending energy that we studied ($k_{\theta}\le3.7\epsilon$).}
Furthermore, the semiflexbile chain still exhibits a 33\% increase in chain size even at low density of $\rho_m=0.014\sigma^{-3}$, indicating that the elongation induced by real stiffness is not significantly affected by interactions with neighboring chains.  

The average number of contact monomers significantly increases as the chain size increases. Specifically, when the chain size reaches 37\%, $\langle n\rangle$ increases by 2.5 times. Notably, Figure \ref{fig:1}d demonstrates the equivalence in the average number of contact monomers when the active polymer and SPC system possess identical end-to-end distances, i.e. at $\Delta T=8 \epsilon/k_B$ and $k_{\theta}=3.7\epsilon$.

\begin{figure}[!t]
    \centering
    \includegraphics[width=0.9\linewidth]{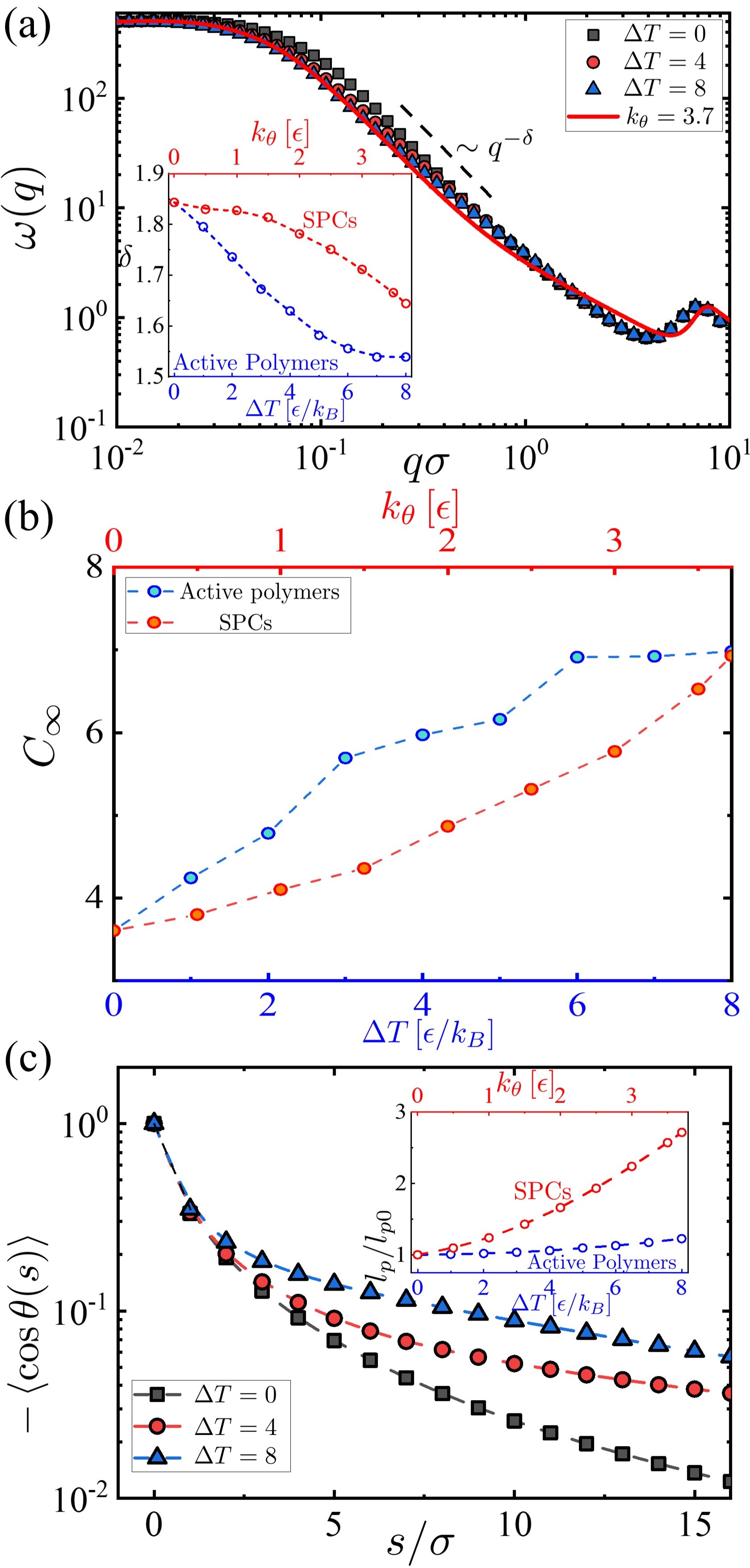}
    \caption{The single-chain structure factors at three different activity differences.  The red line represents $\omega(q)$ for the system of SPCs at bending energy coefficient $k_\theta=3.7\epsilon$. Inset: The scaling exponents between the structure factor and wavevector for the system of active polymers at various activity differences (the blue $x$-axis on the bottom), and the system of SPCs at various bending energy coefficients (the red $x$-axis on the top). (b) The Flory characteristic ratio as a function of activity difference (the blue $x$-axis on the bottom) and bending energy coefficient (the red $x$-axis on the top), respectively.  (c) The bond angle correlation function as a function of contour distance at three activity differences.  Inset: The fitting persistence length reduced by that in $\Delta T$ for the system of active polymers and the system of SPCs. }
    \label{fig:2}
\end{figure}

\subsection{Effective Stiffness}
We further used the single-chain structure factor $\omega(q)$ to characterize the variations in polymer chain conformation resulting from different levels of activity.  Figure \ref{fig:2}a illustrates that these alterations in $\omega(q)$ were more pronounced at an intermediate length of $q\approx 0.1\sigma^{-1}$, corresponding to a real-space length $r=2\pi/q\approx 62.8 \sigma$, which is comparable to the end-to-end  distance $R_e$.   Within this range, the structure factor exhibits a power-law behavior with $\omega\sim q^{-\delta}$. As the activity difference increases to $8\epsilon/k_B$, the exponent $\delta$ decreases from approximately 1.85 to around 1.52 and eventually reaches saturation.  The scaling exponent for a fully flexible Gaussian chain is equal to 2, whereas for a rigid rod it strictly equals to 1.  Therefore, the reduction of the exponent indicates an increased effective stiffness of polymer chains.  Additionally, the system of SPCs also demonstrates a decrease of the scaling exponent in the regime when the bending energy coefficient $k_{\theta}$ increases, as shown in the inset of Figure \ref{fig:2}a.  However, similar to $R_e$ for SPCs, it does not exhibit saturation tendencies even under large $k_{\theta}$.  

We also calculated the Flory characteristic ratio $C_{\infty}$ using the mean-square internal distances, defined as $\langle R^2_e(s)\rangle$ =$\langle R^2_e\rangle(|i-j|,N)$, where $s=|i-j|$ is the distance between $i$-th and $j$th monomers along a chain.  The Flory characteristic ratio $C_{\infty}$ can be readily extracted at large $s$, $\lim_{s\to N}\langle R^2_e(s)\rangle/s=l_b^2C_{\infty}$, quantifying effective chain stiffness.  Figure \ref{fig:2}b shows the characteristic ratio of active polymers increases to saturation at $\Delta T>6\epsilon/k_B$.  The higher $C_\infty$ than those of SPCs systems indicates that chain end activity primarily manifests at the chain size scale rather than the bond length scale.  

At smaller length scale, the effective stiffness of a chain can also be investigated by analyzing the bond angle correlation function, $-\langle cos\theta\rangle$, where $\theta$ represents the internal angle between two bonds in the chain.  As shown in Figure \ref{fig:2}c, the bond angle correlation function decreases with contour distance.  The decay of the function becomes slower for larger differences in activity, suggesting the existence of longer-range correlations.  The initial decay reflects the chain stiffness at the scale of several bond lengths and can be used to extract the persistence length through the fitting with an exponential function $\sim \mathrm{exp}(-sl_p/l_b)$.   The inset of Figure \ref{fig:2}c shows that the persistence length of SPCs with real stiffness is greater than that of active polymers.  Such a more remarkable $l_p$ of SPCs is reasonable considering the fact that the bending force is applied to the local bond angles.  

\begin{figure*}[!t]
    \centering
    \includegraphics[width=0.8\linewidth]{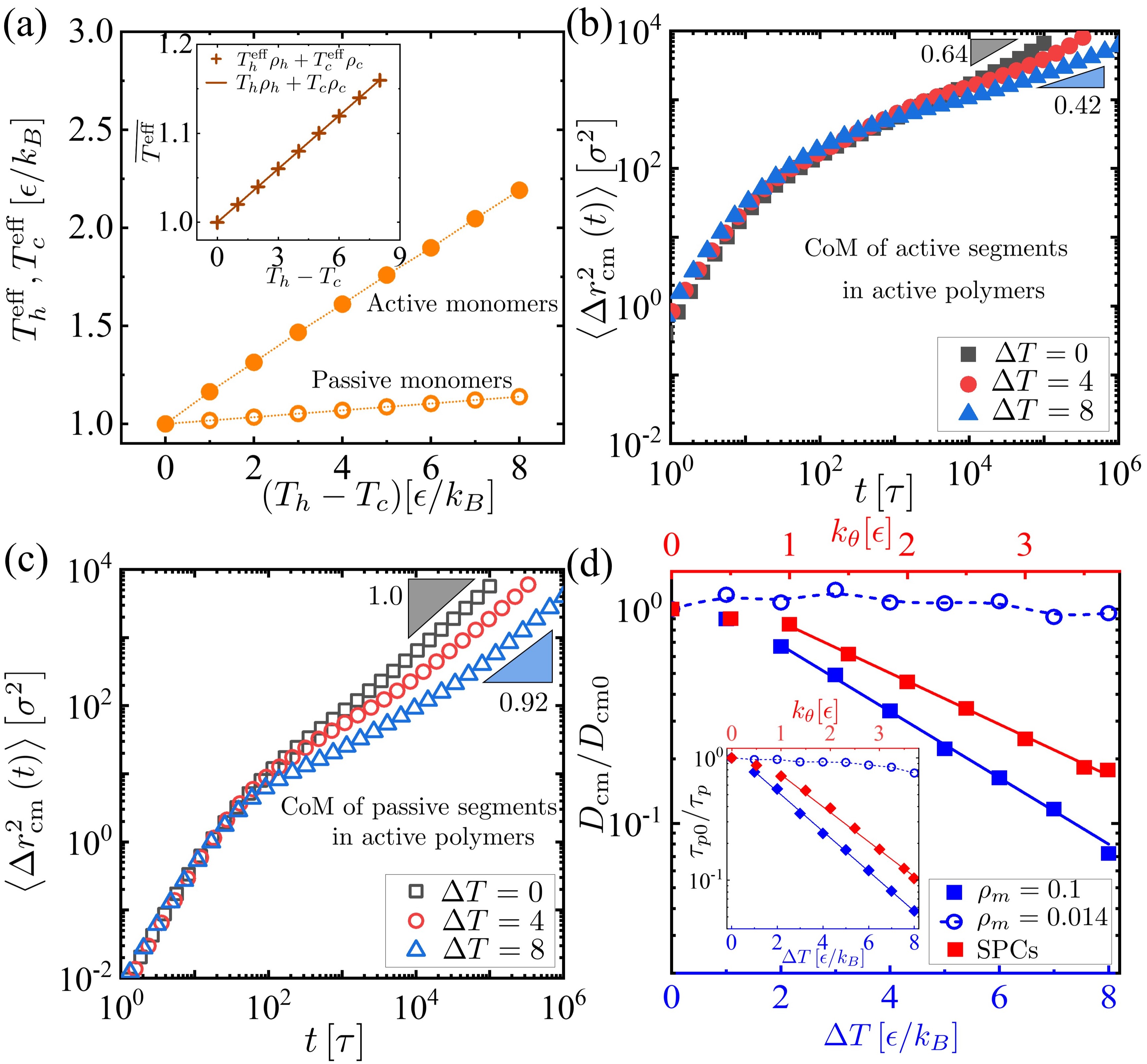}
    \caption{(a) Effective temperature on active and passive monomers in the system of active polymers, respectively.  Inset: the mean effective temperatures calculated by $T_h^{\mathrm{eff}}\rho_h+T_c^{\mathrm{eff}}\rho_c$ (crosses) and $T_h\rho_h+T_c\rho_c$ (line). (b) The mean-squared displacement of the center of mass of active segments on the each end of chain at three different activity differences.  (c) The mean-squared displacement of the center of mass of midchain passive segments.  (d) The center-of-mass diffusion coefficient reduced by that in equilibrium ($\Delta T=0$) as a function of activity difference (the blue $x$-axis on the bottom) and bending energy coefficient(the red $x$-axis on the top), respectively.  The blue dashed line and circles represent the reduced diffusion coefficient in dilution solution with $\rho_m=0.014$.  The solid lines represent the exponential fittings.  }
    \label{fig:3}
\end{figure*}

\subsection{Dynamics and entanglement}
The velocity histograms of active and passive monomers in the system of active polymers are found to satisfy  Maxwell's velocity distributions (see Figure S7 in the SI).  The effective temperature for each type of monomer can be extracted from the distribution according to energy equipartition.  It is evident that the effective temperatures of active and passive monomers exhibit a linear increase in proportion to activity differences as shown in Figure \ref{fig:3}a.   The average effective temperatures for all monomers are calculated as $\overline{T^{\mathrm{eff}}}=T^{\mathrm{eff}}_h\rho_h+T^{\mathrm{eff}}_c\rho_c$ or $\overline{T^{\mathrm{eff}}}=T_h\rho_h+T_c\rho_c$, where $\rho_h$ and $\rho_c$ are the number density of ‘hot’ active monomers and ‘cold’ passive monomers in the system of active polymers, respectively.  It can be seen that those two results coincide as shown in the inset of Figure \ref{fig:3}a.  The average temperature gradually increases with the activity difference.  The observed relationship is consistent with the theoretical prediction based on the modified Einstein relation \cite{Ilker2021}.

Next, the focus of our study is then shifted towards the long-time dynamic behaviors when the average effective temperature increases.  We first calculate the mean-squared displacement (MSD) for the center of mass (CoM) of active terminals located at the ends of polymer chains, $\langle\Delta r_{\mathrm{cm}}^2(t)\rangle_h=\langle |\bm{r}_{\mathrm{cm}}^h(t)-\bm{r}_{\mathrm{cm}}^h(0)|^2\rangle$, and the center-of-mass MSDs for passive monomers situated in the central ‘cold’ region, $\langle\Delta r_{\mathrm{cm}}^2(t)\rangle_c=\langle |\bm{r}_{\mathrm{cm}}^c(t)-\bm{r}_{\mathrm{cm}}^c(0)|^2\rangle$.  Figure \ref{fig:3}b and \ref{fig:3}c show that there is a clear difference in scaling behavior between the active terminal and the midchain passive segment of MSD.  The MSDs for active segments exhibit subdiffusion at different levels of activity differences. {  The Rouse theory predicts a significant subdiffusive behavior of the monomer MSD due to chain connectivity in the limit of long chains, with $MSD\sim t^x$ where $x=0.5$. \cite{Doi1986}  Figure \ref{fig:3}b shows that in an equilibrium system ($\Delta T=0$), this scaling exponent is measured as $x=0.64$.  As the temperature difference increases to $8k_BT$, this exponent further decreases to 0.42, indicating that the influence of chain connectivity become more pronounced as an increase in chain size with temperature difference.  In contrast, the MSDs for midchain passive segments exhibit nearly Fickian diffusion behavior, with a scaling exponent approaching 1 regardless of the activity difference. 
}

The activity-modifying monomers on the terminals surprisingly exhibits distinct anomalous diffusion behaviors compared to those observed in equilibrium polymer physics: a reduction in diffusion is observed at higher ambient temperatures.   The CoM diffusion coefficient of the entire chain is defined as $D_{\mathrm{cm}}=\lim_{t\to\infty}\langle\Delta r_{\mathrm{cm}}^2(t)\rangle/6t$.   Figure \ref{fig:3}d shows that the diffusion coefficient reduced by that at $\Delta T=0$ exponentially decays with the activity difference, $D_r^{\mathrm{act}}=D_{\mathrm{cm}}/D_{\mathrm{cm0}}=D_{\mathrm{cm}}^{\mathrm{act}}(\Delta T)/D_{\mathrm{cm}}^{\mathrm{act}}(\Delta T=0)\sim \mathrm{exp}(-0.36 \Delta T)$.  Similarly, for the SPC system, the CoM diffusion coefficient also follows an exponential decay, $D_{r}^{\mathrm{SPC}}\sim \mathrm{exp}(-0.59k_{\theta})$, albeit with a smaller slope on the semi-log plot.  

Another useful dynamic quantity is the Rouse relaxation time obtained by Rouse mode analysis.  We performed Rouse mode analysis $\langle \bm{X}_p(t) \bm{X}_p(0)\rangle/\langle \bm{X}_p^2\rangle$ to extract chain diffusion coefficients and Rouse relaxation times.  The $p$-th Rouse mode is given by 
\begin{equation}
    \bm{X}_p(t)=\frac{1}{N}\sum_{i=1}^Ncos\Big[\frac{p\pi}{N}\Big(i-\frac{1}{2}\Big)\Big]\bm{r}_i(t)
\end{equation}
where $\bm{r}_i(t)$ is the position vector of bead $i$.  The Rouse mode autocorrelation function can be fitted by a stretched exponential $\langle \bm{X}_p(t) \bm{X}_p(0)\rangle/\langle \bm{X}_p^2\rangle=\mathrm{exp}[(-t/\tau_p)^{\beta_p}]$.  The Rouse mode relaxation is taken as the relaxation of the first normal mode ($p=1$). 

It is interesting that  the inverse of reduced Rouse relaxation times for both active and SPC systems also exhibits exponential decay: $(\tau_{d0}/\tau_{d})^{\mathrm{act}}\sim exp(-0.38 \Delta T)$ and $(\tau_{d0}/\tau_{d})^{\mathrm{SPC}}\sim exp(-0.71 k_{\theta})$.  Notably, the decays of $D_{\mathrm{cm}}$ and $1/\tau_d$ are quite similar in the active system.  

{In an equilibrium system of semidilute passive polymer solution, it is well-known that increased stiffness leads to a higher entanglement number and subsequently results in a decreased diffusion coefficient.\cite{Everaers2008,Milner2020,Wang2003}}
Therefore, we propose that the active terminals in a linear polymer solution can induce effective stiffness, leading to their elongation and entanglement with other chains even in the unentangled semidilute polymer solution, ultimately causing a decrease in their diffusion.   Figure \ref{fig:3}d shows that in low density, even if the chain is stretched, there are fewer interchain interactions, thus, the diffusion coefficient remains almost unchanged.  As a comparison, the real bending energy in the SPC system causes elongation, which enhances interactions with surrounding chains and subsequently also results in the reduction of the diffusion coefficient.  

\begin{figure}[!t]
    \centering
    \includegraphics[width=0.8\linewidth]{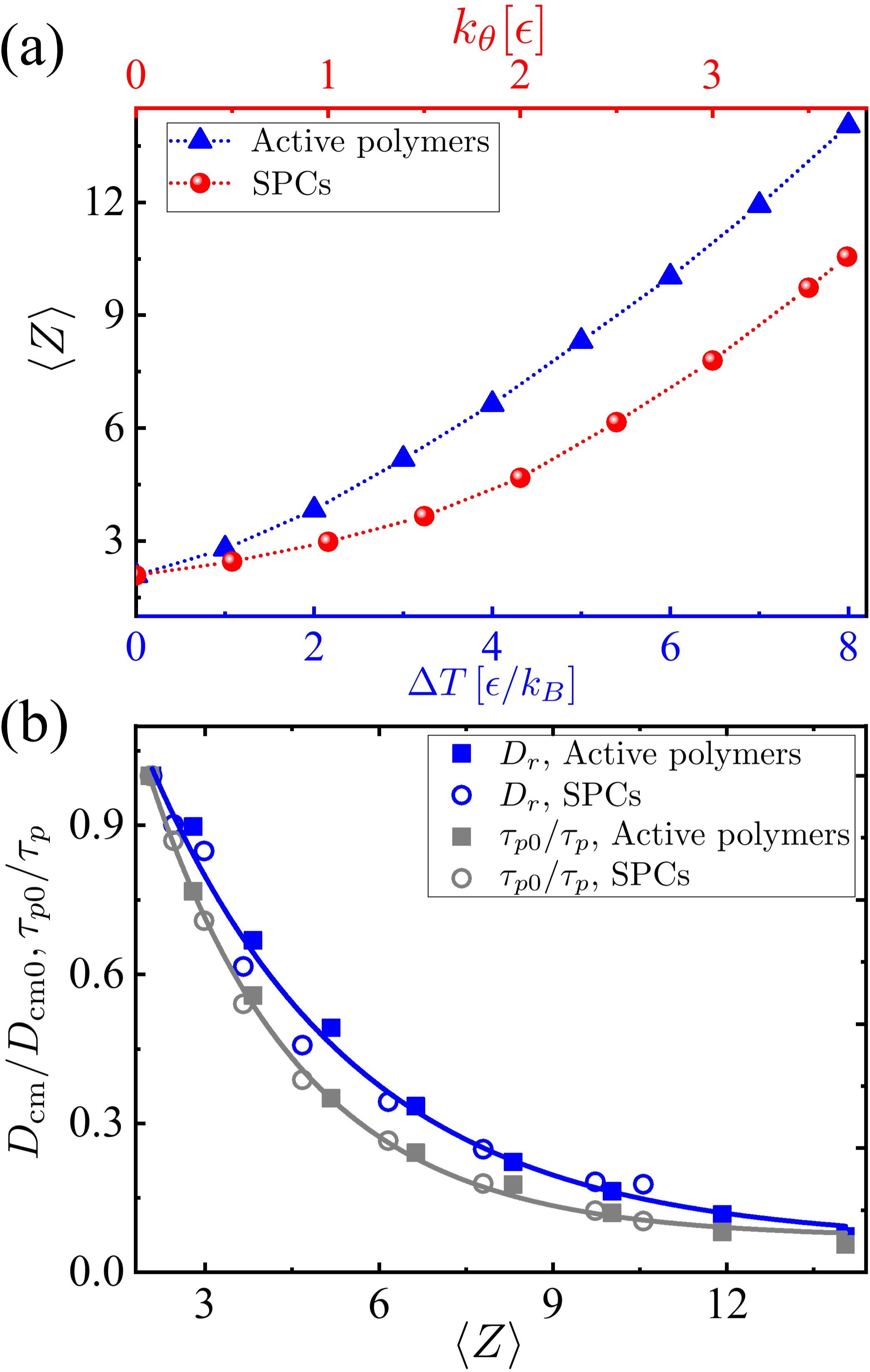}
    \caption{(a) Mean entanglement number as a function of activity difference (active polymers, the blue $x-$axis on the bottom), and bending energy coefficient (SPCs, the red $x-$axis on the top).  (b) The reduced diffusion coefficients are shown in blue diamonds (active polymers) and circles (SPCs), respectively.  On the same $y-$axis, the inverse of reduced Rouse relaxation time are shown in grey diamonds (active polymers) and circles (SPCs).  }
    \label{fig:4}
\end{figure}

We employed the latest Z1 code for the primitive path (PP) analysis to describe entanglement network between chains arising from topological constraints.\cite{Kroger2023}  The entanglement is  characterized by the mean entanglement number per chain $\langle Z\rangle$, entanglement length $N_e$, PP length $L_{\mathrm{pp}}$, and the tube diameter $a_{\mathrm{pp}}$ as listed in Table \ref{Table1} and \ref{Table2}. 

Figure \ref{fig:4}a demonstrates a significant increase in the mean entanglement numbers $\langle Z\rangle$, multiplying by several times from 2.1 to 14, when $\Delta T$ increases to $8\epsilon/k_B$.  The dependence of entanglement number on the bending energy coefficient $k_{\theta}$ is slightly attenuated in SPCs system.  It still remains smaller than the value for active polymers.  $\langle Z\rangle$ for SPC system reaches 10.6 under the same elongation at $k_{\theta}=3.7\epsilon/k_B$. 

{The influence of entanglement on polymer dynamics in semidilute and melt regime has been extensively investigated in previous studies \cite{Halverson2011, Herrmann2012}. } In this study, we present the relationship between diffusion coefficients and Rouse relaxation time with entanglement length in both systems of active polymers and equilibrium SPCs.  The reduced diffusion coefficients of both systems exhibit an exponential decay and collapse onto a master curve, as described by the relation $D_r =1.72\mathrm{exp}(-\langle Z\rangle/3.54)$.  Similarly, the Rouse relaxation times of these two systems as a function of entanglement number also collapse onto a curve, and their inverse follows an exponential decay given by $\tau_{p0}/\tau_p =2.03\mathrm{exp}(-\langle Z\rangle/2.61)$.

A fascinating consequence of these unified curves is that the same entanglement number corresponds to an identical value of a dynamical quantity.  Our simulation reveals that whether it is non-equilibrium active polymers or equilibrium polymers, the topological constraints dominate dynamic properties, such as CoM diffusion and relaxation of the chains, although do not uniquely determine the corresponding structural quantity like chain size and effective stiffness.  

\begin{table*}[t]
\small
\caption{Summary of PP length $L_{\mathrm{pp}}$, tube diameter $a_{\mathrm{pp}}$ and entanglement length $N_e$ for the system of active polymers. }
\begin{tabular*}{1.00\textwidth}{@{\extracolsep{\fill}}llllllllll}
  \hline
  $T_{\textrm{h}}-T_{\textrm{c}}$    &0    &1    &2   &3   &4   &5    &6   &7    &8   \\
  \hline
          $L_{\mathrm{pp}}$                         &52.35      &61.14    &72.34   &86.50   &99.20    &112.66   &124.21     &135.61      &146.91            \\
          $a_{\mathrm{pp}}$                         &31.93      &31.39    &30.65   &29.73   &27.72    &25.94   &24.66     	&23.40       &21.78            \\
          $N_e$             &242.63 	&180.17   &131.13  &96.83 	&75.50 	  &60.22 	&49.96 	    &41.96 	     &35.62             \\
  \hline
  \end{tabular*}
\label{Table1}
\end{table*}

\begin{table*}[t]
\small
\caption{Summary of PP length $L_{\mathrm{pp}}$, tube diameter $a_{\mathrm{pp}}$ and entanglement length $N_e$ for the system of SPCs.}
\begin{tabular*}{1.00\textwidth}{@{\extracolsep{\fill}}llllllllll}
  \hline
  $\textrm{k}_{\theta}$    &0.0    &0.5    &1.0   &1.5   &2.0   &2.5    &3.0   &3.5    &3.7   \\
  \hline
          $L_{\mathrm{pp}}$                         &52.25 	    &56.53 	  &62.79   &69.41 	&80.07 	  &94.28 	&109.49 	&125.73 	 &132.40         \\
          $a_{\mathrm{pp}}$                         &32.41 	    &31.24 	  &30.43   &29.43 	&28.14 	  &26.72 	&25.76 		&24.97	       &24.84         \\
          $N_e$             &248.97 	&204.59   &168.75  &137.56 	&107.35   &81.41 	&64.24 	    &51.46 	     &47.44              \\
  \hline
  \end{tabular*}
\label{Table2}
\end{table*}

\subsection{Active modifications on different sites}
\begin{figure}[!b]
    \centering
    \includegraphics[width=0.8\linewidth]{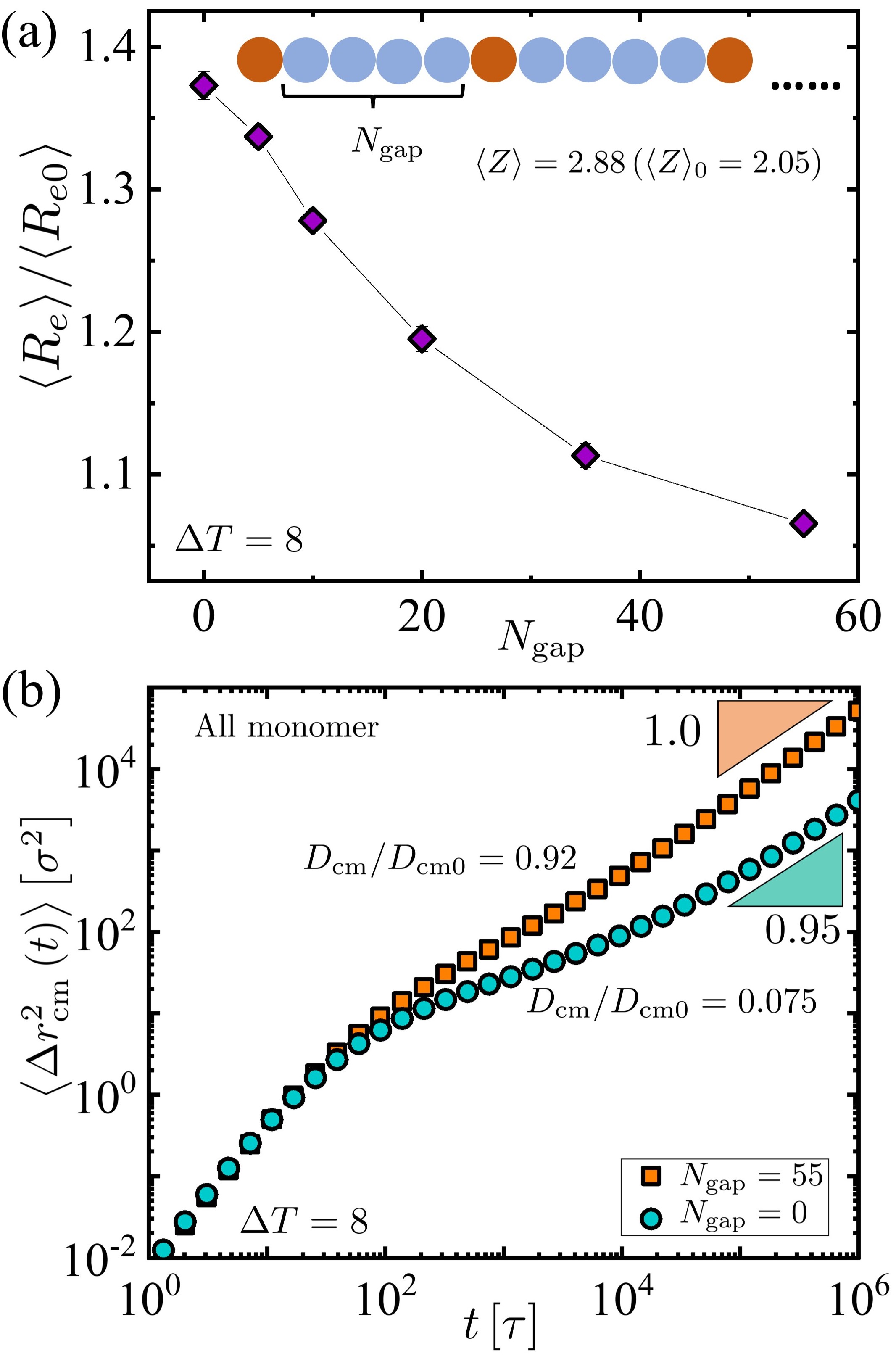}
    \caption{(a) The end-to-end distance as a function of the number of passive monomers between two neighbouring active monomers along a chain.  (b) The CoM mean-square displacement of all monomers in a chain for two different active site distributions, active terminals ($N_{\mathrm{gap}}=0$) and evenly distributed active sites ($N_{\mathrm{gap}}=55$).  }
    \label{fig:5}
\end{figure}

Active modification on different sites along a chain can regulate chain size, dynamics and entanglement.  We considered the influence of different distributions of 10 active monomers.  The number of passive monomers between two active monomers along a chain is set to $N_{\mathrm{gap}}$ as shown in the inset of Figure \ref{fig:5}a.  $N_{\mathrm{gap}}=0$ for active polymers with active terminals, while $N_{\mathrm{gap}}=55$ means that there are approximately even distributions of 10 active monomers along the chain.  

Figure \ref{fig:5}a shows that with the increase of $N_{\mathrm{gap}}$, the reduced end-to-end distance monotonically decreases, ranging from 37\% to 6\%, i.e., the evenly distributed active modification leads to a shorter chain size,  while the corresponding entanglement number does not change significantly ($\langle Z\rangle=2.88$).  In comparison to the system of active ends, the long-time CoM MSD  is increased by more than  an order of magnitude, leading to a higher reduced diffusion, $D_{\mathrm{cm}}/D_{\mathrm{cm0}}\approx 0.92$, despite an observed increase in effective temperature.  

{
\section{Summary}
Our research has revealed that activity triggers chain swelling, effective stiffness, entangled network, and slow dynamics in semidilute unentangled polymer solutions.  We propose that hot active monomers at the chain ends would result in system solidification, deviating significantly from behaviors observed in active small molecules or filaments with real active force.  These findings are significantly inconsistent with previous research results, which indicated that activity can increase fluidity and reduce rigidity. \cite{Aranson2008,Humphrey2002} The paper proposes a novel approach for the regulation of fluid-solid transitions in biological systems through active modification.

The complex relationship between various active self-propulsion mechanisms and their interactions with chain-like structures creates a deeply intricate effect on the activity, structure, entanglement, and dynamics.\cite{Roland2017}  At present, one primary type of active polymers is achieved by exerting active forces on the monomers, directed randomly or along the chain’s contour.  The real active force violates the fluctuation-dissipation theorem in Langevin dynamics.  Randomly oriented activity forces can induce expansion of self-avoiding chains under low Peclet number, while causing swelling at high Peclet number. \cite{Anand2018,Anand2020} Conversely, activity forces along the contour can lead to chain form globule-like conformation and enhance diffusion.  \cite{Bianco2018} Therefore, it is intriguing to investigate and contrast the impacts of activity driven by real forces or hot particles, which should be regarded as a pivotal direction for future research.
}

\begin{acknowledgments}
This work is supported by National Natural Science Foundation of China (No.12374218, 11904320, 11774041, and 12347102), the Fundamental Research Funds for the Central Universities, China under Grant No. SWU-KQ22033 and the Natural Science Foundation of Chongqing, China under Grant No. CSTB2022NSCQ-MSX0512. The authors are grateful to the High Performance Computing Center of Southwest University for carrying out the partial numerical calculations in this work on its blade cluster system .
\end{acknowledgments}

\bibliography{all_acs} 

\begin{thebibliography}{63}%
\makeatletter
\providecommand \@ifxundefined [1]{%
 \@ifx{#1\undefined}
}%
\providecommand \@ifnum [1]{%
 \ifnum #1\expandafter \@firstoftwo
 \else \expandafter \@secondoftwo
 \fi
}%
\providecommand \@ifx [1]{%
 \ifx #1\expandafter \@firstoftwo
 \else \expandafter \@secondoftwo
 \fi
}%
\providecommand \natexlab [1]{#1}%
\providecommand \enquote  [1]{``#1''}%
\providecommand \bibnamefont  [1]{#1}%
\providecommand \bibfnamefont [1]{#1}%
\providecommand \citenamefont [1]{#1}%
\providecommand \href@noop [0]{\@secondoftwo}%
\providecommand \href [0]{\begingroup \@sanitize@url \@href}%
\providecommand \@href[1]{\@@startlink{#1}\@@href}%
\providecommand \@@href[1]{\endgroup#1\@@endlink}%
\providecommand \@sanitize@url [0]{\catcode `\\12\catcode `\$12\catcode
  `\&12\catcode `\#12\catcode `\^12\catcode `\_12\catcode `\%12\relax}%
\providecommand \@@startlink[1]{}%
\providecommand \@@endlink[0]{}%
\providecommand \url  [0]{\begingroup\@sanitize@url \@url }%
\providecommand \@url [1]{\endgroup\@href {#1}{\urlprefix }}%
\providecommand \urlprefix  [0]{URL }%
\providecommand \Eprint [0]{\href }%
\providecommand \doibase [0]{https://doi.org/}%
\providecommand \selectlanguage [0]{\@gobble}%
\providecommand \bibinfo  [0]{\@secondoftwo}%
\providecommand \bibfield  [0]{\@secondoftwo}%
\providecommand \translation [1]{[#1]}%
\providecommand \BibitemOpen [0]{}%
\providecommand \bibitemStop [0]{}%
\providecommand \bibitemNoStop [0]{.\EOS\space}%
\providecommand \EOS [0]{\spacefactor3000\relax}%
\providecommand \BibitemShut  [1]{\csname bibitem#1\endcsname}%
\let\auto@bib@innerbib\@empty
\bibitem [{\citenamefont {Goloborodko}\ \emph {et~al.}(2016)\citenamefont
  {Goloborodko}, \citenamefont {Marko},\ and\ \citenamefont
  {Mirny}}]{Goloborodko2016}%
  \BibitemOpen
  \bibfield  {author} {\bibinfo {author} {\bibfnamefont {A.}~\bibnamefont
  {Goloborodko}}, \bibinfo {author} {\bibfnamefont {J.}~\bibnamefont {Marko}},\
  and\ \bibinfo {author} {\bibfnamefont {L.}~\bibnamefont {Mirny}},\ }\bibfield
   {title} {\bibinfo {title} {Chromosome compaction by active loop extrusion},\
  }\href {https://doi.org/https://doi.org/10.1016/j.bpj.2016.02.041} {\bibfield
   {journal} {\bibinfo  {journal} {Biophys. J.}\ }\textbf {\bibinfo {volume}
  {110}},\ \bibinfo {pages} {2162} (\bibinfo {year} {2016})}\BibitemShut
  {NoStop}%
\bibitem [{\citenamefont {Zhang}\ \emph {et~al.}(2019)\citenamefont {Zhang},
  \citenamefont {Chen}, \citenamefont {Li}, \citenamefont {Taft}, \citenamefont
  {Yao}, \citenamefont {Bai},\ and\ \citenamefont {Xing}}]{Jingyu2019}%
  \BibitemOpen
  \bibfield  {author} {\bibinfo {author} {\bibfnamefont {J.}~\bibnamefont
  {Zhang}}, \bibinfo {author} {\bibfnamefont {H.}~\bibnamefont {Chen}},
  \bibinfo {author} {\bibfnamefont {R.}~\bibnamefont {Li}}, \bibinfo {author}
  {\bibfnamefont {D.~A.}\ \bibnamefont {Taft}}, \bibinfo {author}
  {\bibfnamefont {G.}~\bibnamefont {Yao}}, \bibinfo {author} {\bibfnamefont
  {F.}~\bibnamefont {Bai}},\ and\ \bibinfo {author} {\bibfnamefont
  {J.}~\bibnamefont {Xing}},\ }\bibfield  {title} {\bibinfo {title} {Spatial
  clustering and common regulatory elements correlate with coordinated gene
  expression},\ }\href {https://doi.org/10.1371/journal.pcbi.1006786}
  {\bibfield  {journal} {\bibinfo  {journal} {Plos Comput. Biol.}\ }\textbf
  {\bibinfo {volume} {15}},\ \bibinfo {pages} {1} (\bibinfo {year}
  {2019})}\BibitemShut {NoStop}%
\bibitem [{\citenamefont {Dai}\ and\ \citenamefont {Dai}(2011)}]{Zhiming2011}%
  \BibitemOpen
  \bibfield  {author} {\bibinfo {author} {\bibfnamefont {Z.}~\bibnamefont
  {Dai}}\ and\ \bibinfo {author} {\bibfnamefont {X.}~\bibnamefont {Dai}},\
  }\bibfield  {title} {\bibinfo {title} {{Nuclear colocalization of
  transcription factor target genes strengthens coregulation in yeast}},\
  }\href {https://doi.org/10.1093/nar/gkr689} {\bibfield  {journal} {\bibinfo
  {journal} {Nucleic Acids Res.}\ }\textbf {\bibinfo {volume} {40}},\ \bibinfo
  {pages} {27} (\bibinfo {year} {2011})}\BibitemShut {NoStop}%
\bibitem [{\citenamefont {Bohrer}\ and\ \citenamefont
  {Larson}(2023)}]{Bohrer2023}%
  \BibitemOpen
  \bibfield  {author} {\bibinfo {author} {\bibfnamefont {C.~H.}\ \bibnamefont
  {Bohrer}}\ and\ \bibinfo {author} {\bibfnamefont {D.~R.}\ \bibnamefont
  {Larson}},\ }\bibfield  {title} {\bibinfo {title} {Synthetic analysis of
  chromatin tracing and live-cell imaging indicates pervasive spatial coupling
  between genes},\ }\href {https://doi.org/10.7554/eLife.81861} {\bibfield
  {journal} {\bibinfo  {journal} {eLife}\ }\textbf {\bibinfo {volume} {12}},\
  \bibinfo {pages} {e81861} (\bibinfo {year} {2023})}\BibitemShut {NoStop}%
\bibitem [{\citenamefont {Fanucchi}\ \emph {et~al.}(2013)\citenamefont
  {Fanucchi}, \citenamefont {Shibayama}, \citenamefont {Burd}, \citenamefont
  {Weinberg},\ and\ \citenamefont {Mhlanga}}]{Stephanie2013}%
  \BibitemOpen
  \bibfield  {author} {\bibinfo {author} {\bibfnamefont {S.}~\bibnamefont
  {Fanucchi}}, \bibinfo {author} {\bibfnamefont {Y.}~\bibnamefont {Shibayama}},
  \bibinfo {author} {\bibfnamefont {S.}~\bibnamefont {Burd}}, \bibinfo {author}
  {\bibfnamefont {M.}~\bibnamefont {Weinberg}},\ and\ \bibinfo {author}
  {\bibfnamefont {M.}~\bibnamefont {Mhlanga}},\ }\bibfield  {title} {\bibinfo
  {title} {Chromosomal contact permits transcription between coregulated
  genes},\ }\href {https://doi.org/10.1016/j.cell.2013.09.051} {\bibfield
  {journal} {\bibinfo  {journal} {Cell}\ }\textbf {\bibinfo {volume} {155}},\
  \bibinfo {pages} {606} (\bibinfo {year} {2013})}\BibitemShut {NoStop}%
\bibitem [{\citenamefont {Ganai}\ \emph {et~al.}(2014)\citenamefont {Ganai},
  \citenamefont {Sengupta},\ and\ \citenamefont {Menon}}]{Ganai2014}%
  \BibitemOpen
  \bibfield  {author} {\bibinfo {author} {\bibfnamefont {N.}~\bibnamefont
  {Ganai}}, \bibinfo {author} {\bibfnamefont {S.}~\bibnamefont {Sengupta}},\
  and\ \bibinfo {author} {\bibfnamefont {G.~I.}\ \bibnamefont {Menon}},\
  }\bibfield  {title} {\bibinfo {title} {{Chromosome positioning from
  activity-based segregation}},\ }\href {https://doi.org/10.1093/nar/gkt1417}
  {\bibfield  {journal} {\bibinfo  {journal} {Nucleic Acids Res.}\ }\textbf
  {\bibinfo {volume} {42}},\ \bibinfo {pages} {4145} (\bibinfo {year}
  {2014})}\BibitemShut {NoStop}%
\bibitem [{\citenamefont {Nuebler}\ \emph {et~al.}(2018)\citenamefont
  {Nuebler}, \citenamefont {Fudenberg}, \citenamefont {Imakaev}, \citenamefont
  {Abdennur},\ and\ \citenamefont {Mirny}}]{Nuebler2018}%
  \BibitemOpen
  \bibfield  {author} {\bibinfo {author} {\bibfnamefont {J.}~\bibnamefont
  {Nuebler}}, \bibinfo {author} {\bibfnamefont {G.}~\bibnamefont {Fudenberg}},
  \bibinfo {author} {\bibfnamefont {M.}~\bibnamefont {Imakaev}}, \bibinfo
  {author} {\bibfnamefont {N.}~\bibnamefont {Abdennur}},\ and\ \bibinfo
  {author} {\bibfnamefont {L.~A.}\ \bibnamefont {Mirny}},\ }\bibfield  {title}
  {\bibinfo {title} {Chromatin organization by an interplay of loop extrusion
  and compartmental segregation},\ }\href
  {https://doi.org/10.1073/pnas.1717730115} {\bibfield  {journal} {\bibinfo
  {journal} {P. Natl. Acad. Sci. USA}\ }\textbf {\bibinfo {volume} {115}},\
  \bibinfo {pages} {E6697} (\bibinfo {year} {2018})}\BibitemShut {NoStop}%
\bibitem [{\citenamefont {Mahajan}\ \emph {et~al.}(2022)\citenamefont
  {Mahajan}, \citenamefont {Yan}, \citenamefont {Zidovska}, \citenamefont
  {Saintillan},\ and\ \citenamefont {Shelley}}]{Mahajan2022}%
  \BibitemOpen
  \bibfield  {author} {\bibinfo {author} {\bibfnamefont {A.}~\bibnamefont
  {Mahajan}}, \bibinfo {author} {\bibfnamefont {W.}~\bibnamefont {Yan}},
  \bibinfo {author} {\bibfnamefont {A.}~\bibnamefont {Zidovska}}, \bibinfo
  {author} {\bibfnamefont {D.}~\bibnamefont {Saintillan}},\ and\ \bibinfo
  {author} {\bibfnamefont {M.~J.}\ \bibnamefont {Shelley}},\ }\bibfield
  {title} {\bibinfo {title} {Euchromatin activity enhances segregation and
  compaction of heterochromatin in the cell nucleus},\ }\href
  {https://doi.org/10.1103/PhysRevX.12.041033} {\bibfield  {journal} {\bibinfo
  {journal} {Phys. Rev. X}\ }\textbf {\bibinfo {volume} {12}},\ \bibinfo
  {pages} {041033} (\bibinfo {year} {2022})}\BibitemShut {NoStop}%
\bibitem [{\citenamefont {Saintillan}\ \emph {et~al.}(2018)\citenamefont
  {Saintillan}, \citenamefont {Shelley},\ and\ \citenamefont
  {Zidovska}}]{Saintillan2018}%
  \BibitemOpen
  \bibfield  {author} {\bibinfo {author} {\bibfnamefont {D.}~\bibnamefont
  {Saintillan}}, \bibinfo {author} {\bibfnamefont {M.~J.}\ \bibnamefont
  {Shelley}},\ and\ \bibinfo {author} {\bibfnamefont {A.}~\bibnamefont
  {Zidovska}},\ }\bibfield  {title} {\bibinfo {title} {Extensile motor activity
  drives coherent motions in a model of interphase chromatin},\ }\href
  {https://doi.org/10.1073/pnas.1807073115} {\bibfield  {journal} {\bibinfo
  {journal} {P. Natl. Acad. Sci. USA}\ }\textbf {\bibinfo {volume} {115}},\
  \bibinfo {pages} {11442} (\bibinfo {year} {2018})}\BibitemShut {NoStop}%
\bibitem [{\citenamefont {Zidovska}\ \emph {et~al.}(2013)\citenamefont
  {Zidovska}, \citenamefont {Weitz},\ and\ \citenamefont
  {Mitchison}}]{Zidovska2013}%
  \BibitemOpen
  \bibfield  {author} {\bibinfo {author} {\bibfnamefont {A.}~\bibnamefont
  {Zidovska}}, \bibinfo {author} {\bibfnamefont {D.~A.}\ \bibnamefont
  {Weitz}},\ and\ \bibinfo {author} {\bibfnamefont {T.~J.}\ \bibnamefont
  {Mitchison}},\ }\bibfield  {title} {\bibinfo {title} {Micron-scale coherence
  in interphase chromatin dynamics},\ }\href
  {https://doi.org/10.1073/pnas.1220313110} {\bibfield  {journal} {\bibinfo
  {journal} {P. Natl. Acad. Sci. USA}\ }\textbf {\bibinfo {volume} {110}},\
  \bibinfo {pages} {15555} (\bibinfo {year} {2013})}\BibitemShut {NoStop}%
\bibitem [{\citenamefont {Goychuk}\ \emph {et~al.}(2023)\citenamefont
  {Goychuk}, \citenamefont {Kannan}, \citenamefont {Chakraborty},\ and\
  \citenamefont {Kardar}}]{Goychuk2023}%
  \BibitemOpen
  \bibfield  {author} {\bibinfo {author} {\bibfnamefont {A.}~\bibnamefont
  {Goychuk}}, \bibinfo {author} {\bibfnamefont {D.}~\bibnamefont {Kannan}},
  \bibinfo {author} {\bibfnamefont {A.~K.}\ \bibnamefont {Chakraborty}},\ and\
  \bibinfo {author} {\bibfnamefont {M.}~\bibnamefont {Kardar}},\ }\bibfield
  {title} {\bibinfo {title} {Polymer folding through active processes recreates
  features of genome organization},\ }\href
  {https://doi.org/10.1073/pnas.2221726120} {\bibfield  {journal} {\bibinfo
  {journal} {P. Natl. Acad. Sci. USA}\ }\textbf {\bibinfo {volume} {120}},\
  \bibinfo {pages} {e2221726120} (\bibinfo {year} {2023})}\BibitemShut
  {NoStop}%
\bibitem [{\citenamefont {Misteli}(2020)}]{Misteli2020}%
  \BibitemOpen
  \bibfield  {author} {\bibinfo {author} {\bibfnamefont {T.}~\bibnamefont
  {Misteli}},\ }\bibfield  {title} {\bibinfo {title} {The self-organizing
  genome: Principles of genome architecture and function},\ }\href
  {https://doi.org/https://doi.org/10.1016/j.cell.2020.09.014} {\bibfield
  {journal} {\bibinfo  {journal} {Cell}\ }\textbf {\bibinfo {volume} {183}},\
  \bibinfo {pages} {28} (\bibinfo {year} {2020})}\BibitemShut {NoStop}%
\bibitem [{\citenamefont {Biswas}\ \emph {et~al.}(2017)\citenamefont {Biswas},
  \citenamefont {Manna}, \citenamefont {Laskar}, \citenamefont {Kumar},
  \citenamefont {Adhikari},\ and\ \citenamefont {Kumaraswamy}}]{Biswas2017}%
  \BibitemOpen
  \bibfield  {author} {\bibinfo {author} {\bibfnamefont {B.}~\bibnamefont
  {Biswas}}, \bibinfo {author} {\bibfnamefont {R.~K.}\ \bibnamefont {Manna}},
  \bibinfo {author} {\bibfnamefont {A.}~\bibnamefont {Laskar}}, \bibinfo
  {author} {\bibfnamefont {P.~B.~S.}\ \bibnamefont {Kumar}}, \bibinfo {author}
  {\bibfnamefont {R.}~\bibnamefont {Adhikari}},\ and\ \bibinfo {author}
  {\bibfnamefont {G.}~\bibnamefont {Kumaraswamy}},\ }\bibfield  {title}
  {\bibinfo {title} {Linking catalyst-coated isotropic colloids into
  “active” flexible chains enhances their diffusivity},\ }\href
  {https://doi.org/10.1021/acsnano.7b04265} {\bibfield  {journal} {\bibinfo
  {journal} {ACS Nano}\ }\textbf {\bibinfo {volume} {11}},\ \bibinfo {pages}
  {10025} (\bibinfo {year} {2017})}\BibitemShut {NoStop}%
\bibitem [{\citenamefont {Di~Leonardo}(2016)}]{Leonardo2016}%
  \BibitemOpen
  \bibfield  {author} {\bibinfo {author} {\bibfnamefont {R.}~\bibnamefont
  {Di~Leonardo}},\ }\bibfield  {title} {\bibinfo {title} {Controlled collective
  motions},\ }\href {https://doi.org/10.1038/nmat4761} {\bibfield  {journal}
  {\bibinfo  {journal} {Nat. Mater.}\ }\textbf {\bibinfo {volume} {15}},\
  \bibinfo {pages} {1057} (\bibinfo {year} {2016})}\BibitemShut {NoStop}%
\bibitem [{\citenamefont {Fan}\ and\ \citenamefont
  {Walther}(2022)}]{Xinlong2022}%
  \BibitemOpen
  \bibfield  {author} {\bibinfo {author} {\bibfnamefont {X.}~\bibnamefont
  {Fan}}\ and\ \bibinfo {author} {\bibfnamefont {A.}~\bibnamefont {Walther}},\
  }\bibfield  {title} {\bibinfo {title} {1d colloidal chains: recent progress
  from formation to emergent properties and applications},\ }\href
  {https://doi.org/10.1039/D2CS00112H} {\bibfield  {journal} {\bibinfo
  {journal} {Chem. Soc. Rev.}\ }\textbf {\bibinfo {volume} {51}},\ \bibinfo
  {pages} {4023} (\bibinfo {year} {2022})}\BibitemShut {NoStop}%
\bibitem [{\citenamefont {Yan}\ \emph {et~al.}(2016)\citenamefont {Yan},
  \citenamefont {Han}, \citenamefont {Zhang}, \citenamefont {Xu}, \citenamefont
  {Luijten},\ and\ \citenamefont {Granick}}]{Yan2016}%
  \BibitemOpen
  \bibfield  {author} {\bibinfo {author} {\bibfnamefont {J.}~\bibnamefont
  {Yan}}, \bibinfo {author} {\bibfnamefont {M.}~\bibnamefont {Han}}, \bibinfo
  {author} {\bibfnamefont {J.}~\bibnamefont {Zhang}}, \bibinfo {author}
  {\bibfnamefont {C.}~\bibnamefont {Xu}}, \bibinfo {author} {\bibfnamefont
  {E.}~\bibnamefont {Luijten}},\ and\ \bibinfo {author} {\bibfnamefont
  {S.}~\bibnamefont {Granick}},\ }\bibfield  {title} {\bibinfo {title}
  {Reconfiguring active particles by electrostatic imbalance},\ }\href
  {https://doi.org/10.1038/nmat4696} {\bibfield  {journal} {\bibinfo  {journal}
  {Nat. Mater.}\ }\textbf {\bibinfo {volume} {15}},\ \bibinfo {pages} {1095}
  (\bibinfo {year} {2016})}\BibitemShut {NoStop}%
\bibitem [{\citenamefont {Palagi}\ and\ \citenamefont
  {Fischer}(2018)}]{Stefano2018}%
  \BibitemOpen
  \bibfield  {author} {\bibinfo {author} {\bibfnamefont {S.}~\bibnamefont
  {Palagi}}\ and\ \bibinfo {author} {\bibfnamefont {P.}~\bibnamefont
  {Fischer}},\ }\bibfield  {title} {\bibinfo {title} {Bioinspired
  microrobots},\ }\href {https://doi.org/10.1038/s41578-018-0016-9} {\bibfield
  {journal} {\bibinfo  {journal} {Nat. Rev. Mater.}\ }\textbf {\bibinfo
  {volume} {3}},\ \bibinfo {pages} {113} (\bibinfo {year} {2018})}\BibitemShut
  {NoStop}%
\bibitem [{\citenamefont {Park}\ \emph {et~al.}(2017)\citenamefont {Park},
  \citenamefont {Zhuang}, \citenamefont {Yasa},\ and\ \citenamefont
  {Sitti}}]{Park2017}%
  \BibitemOpen
  \bibfield  {author} {\bibinfo {author} {\bibfnamefont {B.-W.}\ \bibnamefont
  {Park}}, \bibinfo {author} {\bibfnamefont {J.}~\bibnamefont {Zhuang}},
  \bibinfo {author} {\bibfnamefont {O.}~\bibnamefont {Yasa}},\ and\ \bibinfo
  {author} {\bibfnamefont {M.}~\bibnamefont {Sitti}},\ }\bibfield  {title}
  {\bibinfo {title} {Multifunctional bacteria-driven microswimmers for targeted
  active drug delivery},\ }\href {https://doi.org/10.1021/acsnano.7b03207}
  {\bibfield  {journal} {\bibinfo  {journal} {ACS Nano}\ }\textbf {\bibinfo
  {volume} {11}},\ \bibinfo {pages} {8910} (\bibinfo {year}
  {2017})}\BibitemShut {NoStop}%
\bibitem [{\citenamefont {Ober}\ \emph {et~al.}(2015)\citenamefont {Ober},
  \citenamefont {Foresti},\ and\ \citenamefont {Lewis}}]{Ober2015}%
  \BibitemOpen
  \bibfield  {author} {\bibinfo {author} {\bibfnamefont {T.~J.}\ \bibnamefont
  {Ober}}, \bibinfo {author} {\bibfnamefont {D.}~\bibnamefont {Foresti}},\ and\
  \bibinfo {author} {\bibfnamefont {J.~A.}\ \bibnamefont {Lewis}},\ }\bibfield
  {title} {\bibinfo {title} {Active mixing of complex fluids at the
  microscale},\ }\href@noop {} {\bibfield  {journal} {\bibinfo  {journal} {P.
  Natl. Acad. Sci. USA}\ }\textbf {\bibinfo {volume} {112}},\ \bibinfo {pages}
  {12293} (\bibinfo {year} {2015})}\BibitemShut {NoStop}%
\bibitem [{\citenamefont {Gao}\ \emph {et~al.}(2019)\citenamefont {Gao},
  \citenamefont {Oh}, \citenamefont {Tu}, \citenamefont {Chang},\ and\
  \citenamefont {Li}}]{Gao2019}%
  \BibitemOpen
  \bibfield  {author} {\bibinfo {author} {\bibfnamefont {L.}~\bibnamefont
  {Gao}}, \bibinfo {author} {\bibfnamefont {J.}~\bibnamefont {Oh}}, \bibinfo
  {author} {\bibfnamefont {Y.}~\bibnamefont {Tu}}, \bibinfo {author}
  {\bibfnamefont {T.}~\bibnamefont {Chang}},\ and\ \bibinfo {author}
  {\bibfnamefont {C.~Y.}\ \bibnamefont {Li}},\ }\bibfield  {title} {\bibinfo
  {title} {Glass transition temperature of cyclic polystyrene and the linear
  counterpart contamination effect},\ }\href
  {https://doi.org/https://doi.org/10.1016/j.polymer.2019.03.018} {\bibfield
  {journal} {\bibinfo  {journal} {Polymer}\ }\textbf {\bibinfo {volume}
  {170}},\ \bibinfo {pages} {198} (\bibinfo {year} {2019})}\BibitemShut
  {NoStop}%
\bibitem [{\citenamefont {Pipertzis}\ \emph {et~al.}(2018)\citenamefont
  {Pipertzis}, \citenamefont {Hossain}, \citenamefont {Monteiro},\ and\
  \citenamefont {Floudas}}]{George2018}%
  \BibitemOpen
  \bibfield  {author} {\bibinfo {author} {\bibfnamefont {A.}~\bibnamefont
  {Pipertzis}}, \bibinfo {author} {\bibfnamefont {M.~D.}\ \bibnamefont
  {Hossain}}, \bibinfo {author} {\bibfnamefont {M.~J.}\ \bibnamefont
  {Monteiro}},\ and\ \bibinfo {author} {\bibfnamefont {G.}~\bibnamefont
  {Floudas}},\ }\bibfield  {title} {\bibinfo {title} {Segmental dynamics in
  multicyclic polystyrenes},\ }\href
  {https://doi.org/10.1021/acs.macromol.7b02579} {\bibfield  {journal}
  {\bibinfo  {journal} {Macromolecules}\ }\textbf {\bibinfo {volume} {51}},\
  \bibinfo {pages} {1488} (\bibinfo {year} {2018})}\BibitemShut {NoStop}%
\bibitem [{\citenamefont {Smrek}\ \emph {et~al.}(2020)\citenamefont {Smrek},
  \citenamefont {Chubak}, \citenamefont {Likos},\ and\ \citenamefont
  {Kremer}}]{Smrek2020}%
  \BibitemOpen
  \bibfield  {author} {\bibinfo {author} {\bibfnamefont {J.}~\bibnamefont
  {Smrek}}, \bibinfo {author} {\bibfnamefont {I.}~\bibnamefont {Chubak}},
  \bibinfo {author} {\bibfnamefont {C.~N.}\ \bibnamefont {Likos}},\ and\
  \bibinfo {author} {\bibfnamefont {K.}~\bibnamefont {Kremer}},\ }\bibfield
  {title} {\bibinfo {title} {Active topological glass},\ }\href@noop {}
  {\bibfield  {journal} {\bibinfo  {journal} {Nat. Commun.}\ }\textbf {\bibinfo
  {volume} {11}} (\bibinfo {year} {2020})}\BibitemShut {NoStop}%
\bibitem [{\citenamefont {Ubertini}\ \emph {et~al.}(2022)\citenamefont
  {Ubertini}, \citenamefont {Smrek},\ and\ \citenamefont
  {Rosa}}]{Ubertini2022}%
  \BibitemOpen
  \bibfield  {author} {\bibinfo {author} {\bibfnamefont {M.~A.}\ \bibnamefont
  {Ubertini}}, \bibinfo {author} {\bibfnamefont {J.}~\bibnamefont {Smrek}},\
  and\ \bibinfo {author} {\bibfnamefont {A.}~\bibnamefont {Rosa}},\ }\bibfield
  {title} {\bibinfo {title} {Entanglement length scale separates threading from
  branching of unknotted and non-concatenated ring polymers in melts},\ }\href
  {https://doi.org/10.1021/acs.macromol.2c01264} {\bibfield  {journal}
  {\bibinfo  {journal} {Macromolecules}\ }\textbf {\bibinfo {volume} {55}},\
  \bibinfo {pages} {10723} (\bibinfo {year} {2022})}\BibitemShut {NoStop}%
\bibitem [{\citenamefont {Micheletti}\ \emph {et~al.}(2024)\citenamefont
  {Micheletti}, \citenamefont {Chubak}, \citenamefont {Orlandini},\ and\
  \citenamefont {Smrek}}]{Micheletti2024}%
  \BibitemOpen
  \bibfield  {author} {\bibinfo {author} {\bibfnamefont {C.}~\bibnamefont
  {Micheletti}}, \bibinfo {author} {\bibfnamefont {I.}~\bibnamefont {Chubak}},
  \bibinfo {author} {\bibfnamefont {E.}~\bibnamefont {Orlandini}},\ and\
  \bibinfo {author} {\bibfnamefont {J.}~\bibnamefont {Smrek}},\ }\bibfield
  {title} {\bibinfo {title} {Topology-based detection and tracking of deadlocks
  reveal aging of active ring melts},\ }\href
  {https://doi.org/10.1021/acsmacrolett.3c00567} {\bibfield  {journal}
  {\bibinfo  {journal} {ACS Macro Lett.}\ }\textbf {\bibinfo {volume} {13}},\
  \bibinfo {pages} {124} (\bibinfo {year} {2024})}\BibitemShut {NoStop}%
\bibitem [{\citenamefont {Smrek}\ and\ \citenamefont
  {Kremer}(2017)}]{Smrek2017PRL}%
  \BibitemOpen
  \bibfield  {author} {\bibinfo {author} {\bibfnamefont {J.}~\bibnamefont
  {Smrek}}\ and\ \bibinfo {author} {\bibfnamefont {K.}~\bibnamefont {Kremer}},\
  }\bibfield  {title} {\bibinfo {title} {Small activity differences drive phase
  separation in active-passive polymer mixtures},\ }\href
  {https://doi.org/10.1103/PhysRevLett.118.098002} {\bibfield  {journal}
  {\bibinfo  {journal} {Phys. Rev. Lett.}\ }\textbf {\bibinfo {volume} {118}},\
  \bibinfo {pages} {098002} (\bibinfo {year} {2017})}\BibitemShut {NoStop}%
\bibitem [{\citenamefont {Bruinsma}\ \emph {et~al.}(2014)\citenamefont
  {Bruinsma}, \citenamefont {Grosberg}, \citenamefont {Rabin},\ and\
  \citenamefont {Zidovska}}]{Bruinsma2014}%
  \BibitemOpen
  \bibfield  {author} {\bibinfo {author} {\bibfnamefont {R.}~\bibnamefont
  {Bruinsma}}, \bibinfo {author} {\bibfnamefont {A.}~\bibnamefont {Grosberg}},
  \bibinfo {author} {\bibfnamefont {Y.}~\bibnamefont {Rabin}},\ and\ \bibinfo
  {author} {\bibfnamefont {A.}~\bibnamefont {Zidovska}},\ }\bibfield  {title}
  {\bibinfo {title} {Chromatin hydrodynamics},\ }\href
  {https://doi.org/https://doi.org/10.1016/j.bpj.2014.03.038} {\bibfield
  {journal} {\bibinfo  {journal} {Biophys. J.}\ }\textbf {\bibinfo {volume}
  {106}},\ \bibinfo {pages} {1871} (\bibinfo {year} {2014})}\BibitemShut
  {NoStop}%
\bibitem [{\citenamefont {Li}\ \emph {et~al.}(2023)\citenamefont {Li},
  \citenamefont {Wu}, \citenamefont {Hao}, \citenamefont {Lei},\ and\
  \citenamefont {Ma}}]{Li2023}%
  \BibitemOpen
  \bibfield  {author} {\bibinfo {author} {\bibfnamefont {J.-X.}\ \bibnamefont
  {Li}}, \bibinfo {author} {\bibfnamefont {S.}~\bibnamefont {Wu}}, \bibinfo
  {author} {\bibfnamefont {L.-L.}\ \bibnamefont {Hao}}, \bibinfo {author}
  {\bibfnamefont {Q.-L.}\ \bibnamefont {Lei}},\ and\ \bibinfo {author}
  {\bibfnamefont {Y.-Q.}\ \bibnamefont {Ma}},\ }\bibfield  {title} {\bibinfo
  {title} {Nonequilibrium structural and dynamic behaviors of polar active
  polymer controlled by head activity},\ }\href
  {https://doi.org/10.1103/PhysRevResearch.5.043064} {\bibfield  {journal}
  {\bibinfo  {journal} {Phys. Rev. Res.}\ }\textbf {\bibinfo {volume} {5}},\
  \bibinfo {pages} {043064} (\bibinfo {year} {2023})}\BibitemShut {NoStop}%
\bibitem [{\citenamefont {Anand}\ and\ \citenamefont
  {Singh}(2018)}]{Anand2018}%
  \BibitemOpen
  \bibfield  {author} {\bibinfo {author} {\bibfnamefont {S.~K.}\ \bibnamefont
  {Anand}}\ and\ \bibinfo {author} {\bibfnamefont {S.~P.}\ \bibnamefont
  {Singh}},\ }\bibfield  {title} {\bibinfo {title} {Structure and dynamics of a
  self-propelled semiflexible filament},\ }\href
  {https://doi.org/10.1103/PhysRevE.98.042501} {\bibfield  {journal} {\bibinfo
  {journal} {Phys. Rev. E}\ }\textbf {\bibinfo {volume} {98}},\ \bibinfo
  {pages} {042501} (\bibinfo {year} {2018})}\BibitemShut {NoStop}%
\bibitem [{\citenamefont {Isele-Holder}\ \emph {et~al.}(2015)\citenamefont
  {Isele-Holder}, \citenamefont {Elgeti},\ and\ \citenamefont
  {Gompper}}]{Rolf2015}%
  \BibitemOpen
  \bibfield  {author} {\bibinfo {author} {\bibfnamefont {R.~E.}\ \bibnamefont
  {Isele-Holder}}, \bibinfo {author} {\bibfnamefont {J.}~\bibnamefont
  {Elgeti}},\ and\ \bibinfo {author} {\bibfnamefont {G.}~\bibnamefont
  {Gompper}},\ }\bibfield  {title} {\bibinfo {title} {Self-propelled worm-like
  filaments: spontaneous spiral formation, structure, and dynamics},\
  }\href@noop {} {\bibfield  {journal} {\bibinfo  {journal} {Soft matter}\
  }\textbf {\bibinfo {volume} {11}},\ \bibinfo {pages} {7181} (\bibinfo {year}
  {2015})}\BibitemShut {NoStop}%
\bibitem [{\citenamefont {Isele-Holder}\ \emph {et~al.}(2016)\citenamefont
  {Isele-Holder}, \citenamefont {J{\"a}ger}, \citenamefont {Saggiorato},
  \citenamefont {Elgeti},\ and\ \citenamefont {Gompper}}]{Rolf2016}%
  \BibitemOpen
  \bibfield  {author} {\bibinfo {author} {\bibfnamefont {R.~E.}\ \bibnamefont
  {Isele-Holder}}, \bibinfo {author} {\bibfnamefont {J.}~\bibnamefont
  {J{\"a}ger}}, \bibinfo {author} {\bibfnamefont {G.}~\bibnamefont
  {Saggiorato}}, \bibinfo {author} {\bibfnamefont {J.}~\bibnamefont {Elgeti}},\
  and\ \bibinfo {author} {\bibfnamefont {G.}~\bibnamefont {Gompper}},\
  }\bibfield  {title} {\bibinfo {title} {Dynamics of self-propelled filaments
  pushing a load},\ }\href@noop {} {\bibfield  {journal} {\bibinfo  {journal}
  {Soft Matter}\ }\textbf {\bibinfo {volume} {12}},\ \bibinfo {pages} {8495}
  (\bibinfo {year} {2016})}\BibitemShut {NoStop}%
\bibitem [{\citenamefont {Jain}\ and\ \citenamefont {Thakur}(2022)}]{Jain2022}%
  \BibitemOpen
  \bibfield  {author} {\bibinfo {author} {\bibfnamefont {N.}~\bibnamefont
  {Jain}}\ and\ \bibinfo {author} {\bibfnamefont {S.}~\bibnamefont {Thakur}},\
  }\bibfield  {title} {\bibinfo {title} {Collapse dynamics of chemically active
  flexible polymer},\ }\href {https://doi.org/10.1021/acs.macromol.1c02502}
  {\bibfield  {journal} {\bibinfo  {journal} {Macromolecules}\ }\textbf
  {\bibinfo {volume} {55}},\ \bibinfo {pages} {2375} (\bibinfo {year}
  {2022})}\BibitemShut {NoStop}%
\bibitem [{\citenamefont {Natali}\ \emph {et~al.}(2020)\citenamefont {Natali},
  \citenamefont {Caprini},\ and\ \citenamefont {Cecconi}}]{Natali2020}%
  \BibitemOpen
  \bibfield  {author} {\bibinfo {author} {\bibfnamefont {L.}~\bibnamefont
  {Natali}}, \bibinfo {author} {\bibfnamefont {L.}~\bibnamefont {Caprini}},\
  and\ \bibinfo {author} {\bibfnamefont {F.}~\bibnamefont {Cecconi}},\
  }\bibfield  {title} {\bibinfo {title} {How a local active force modifies the
  structural properties of polymers},\ }\href
  {https://doi.org/10.1039/C9SM02258A} {\bibfield  {journal} {\bibinfo
  {journal} {Soft Matter}\ }\textbf {\bibinfo {volume} {16}},\ \bibinfo {pages}
  {2594} (\bibinfo {year} {2020})}\BibitemShut {NoStop}%
\bibitem [{\citenamefont {Bianco}\ \emph {et~al.}(2018)\citenamefont {Bianco},
  \citenamefont {Locatelli},\ and\ \citenamefont {Malgaretti}}]{Bianco2018}%
  \BibitemOpen
  \bibfield  {author} {\bibinfo {author} {\bibfnamefont {V.}~\bibnamefont
  {Bianco}}, \bibinfo {author} {\bibfnamefont {E.}~\bibnamefont {Locatelli}},\
  and\ \bibinfo {author} {\bibfnamefont {P.}~\bibnamefont {Malgaretti}},\
  }\bibfield  {title} {\bibinfo {title} {Globulelike conformation and enhanced
  diffusion of active polymers},\ }\href
  {https://doi.org/10.1103/PhysRevLett.121.217802} {\bibfield  {journal}
  {\bibinfo  {journal} {Phys. Rev. Lett.}\ }\textbf {\bibinfo {volume} {121}},\
  \bibinfo {pages} {217802} (\bibinfo {year} {2018})}\BibitemShut {NoStop}%
\bibitem [{\citenamefont {Jiao}\ \emph {et~al.}(2023)\citenamefont {Jiao},
  \citenamefont {Wang}, \citenamefont {Tian},\ and\ \citenamefont
  {Chen}}]{Yang2023}%
  \BibitemOpen
  \bibfield  {author} {\bibinfo {author} {\bibfnamefont {Y.}~\bibnamefont
  {Jiao}}, \bibinfo {author} {\bibfnamefont {J.}~\bibnamefont {Wang}}, \bibinfo
  {author} {\bibfnamefont {W.-d.}\ \bibnamefont {Tian}},\ and\ \bibinfo
  {author} {\bibfnamefont {K.}~\bibnamefont {Chen}},\ }\bibfield  {title}
  {\bibinfo {title} {Configuration and dynamics of a self-propelled diblock
  copolymer chain},\ }\href {https://doi.org/10.1039/D3SM00596H} {\bibfield
  {journal} {\bibinfo  {journal} {Soft Matter}\ }\textbf {\bibinfo {volume}
  {19}},\ \bibinfo {pages} {5468} (\bibinfo {year} {2023})}\BibitemShut
  {NoStop}%
\bibitem [{\citenamefont {Yan}\ \emph {et~al.}(2023)\citenamefont {Yan},
  \citenamefont {Tan}, \citenamefont {Wang},\ and\ \citenamefont
  {Zhao}}]{Yan2023}%
  \BibitemOpen
  \bibfield  {author} {\bibinfo {author} {\bibfnamefont {R.}~\bibnamefont
  {Yan}}, \bibinfo {author} {\bibfnamefont {F.}~\bibnamefont {Tan}}, \bibinfo
  {author} {\bibfnamefont {J.}~\bibnamefont {Wang}},\ and\ \bibinfo {author}
  {\bibfnamefont {N.}~\bibnamefont {Zhao}},\ }\bibfield  {title} {\bibinfo
  {title} {Conformation and dynamics of an active filament in crowded media},\
  }\href@noop {} {\bibfield  {journal} {\bibinfo  {journal} {J. Chem. Phys.}\
  }\textbf {\bibinfo {volume} {158}} (\bibinfo {year} {2023})}\BibitemShut
  {NoStop}%
\bibitem [{\citenamefont {Loi}\ \emph {et~al.}(2011{\natexlab{a}})\citenamefont
  {Loi}, \citenamefont {Mossa},\ and\ \citenamefont
  {Cugliandolo}}]{Davide2011}%
  \BibitemOpen
  \bibfield  {author} {\bibinfo {author} {\bibfnamefont {D.}~\bibnamefont
  {Loi}}, \bibinfo {author} {\bibfnamefont {S.}~\bibnamefont {Mossa}},\ and\
  \bibinfo {author} {\bibfnamefont {L.~F.}\ \bibnamefont {Cugliandolo}},\
  }\bibfield  {title} {\bibinfo {title} {Effective temperature of active
  complex matter},\ }\href {https://doi.org/10.1039/C0SM01484B} {\bibfield
  {journal} {\bibinfo  {journal} {Soft Matter}\ }\textbf {\bibinfo {volume}
  {7}},\ \bibinfo {pages} {3726} (\bibinfo {year}
  {2011}{\natexlab{a}})}\BibitemShut {NoStop}%
\bibitem [{\citenamefont {Loi}\ \emph {et~al.}(2011{\natexlab{b}})\citenamefont
  {Loi}, \citenamefont {Mossa},\ and\ \citenamefont
  {Cugliandolo}}]{Davide201102}%
  \BibitemOpen
  \bibfield  {author} {\bibinfo {author} {\bibfnamefont {D.}~\bibnamefont
  {Loi}}, \bibinfo {author} {\bibfnamefont {S.}~\bibnamefont {Mossa}},\ and\
  \bibinfo {author} {\bibfnamefont {L.~F.}\ \bibnamefont {Cugliandolo}},\
  }\bibfield  {title} {\bibinfo {title} {Non-conservative forces and effective
  temperatures in active polymers},\ }\href
  {https://doi.org/10.1039/C1SM05819C} {\bibfield  {journal} {\bibinfo
  {journal} {Soft Matter}\ }\textbf {\bibinfo {volume} {7}},\ \bibinfo {pages}
  {10193} (\bibinfo {year} {2011}{\natexlab{b}})}\BibitemShut {NoStop}%
\bibitem [{\citenamefont {Vatin}\ \emph {et~al.}(2024)\citenamefont {Vatin},
  \citenamefont {Kundu},\ and\ \citenamefont {Locatelli}}]{Locatelli2024}%
  \BibitemOpen
  \bibfield  {author} {\bibinfo {author} {\bibfnamefont {M.}~\bibnamefont
  {Vatin}}, \bibinfo {author} {\bibfnamefont {S.}~\bibnamefont {Kundu}},\ and\
  \bibinfo {author} {\bibfnamefont {E.}~\bibnamefont {Locatelli}},\ }\bibfield
  {title} {\bibinfo {title} {Conformation and dynamics of partially active
  linear polymers},\ }\href {https://doi.org/10.1039/D3SM01162C} {\bibfield
  {journal} {\bibinfo  {journal} {Soft Matter}\ }\textbf {\bibinfo {volume}
  {20}},\ \bibinfo {pages} {1892} (\bibinfo {year} {2024})}\BibitemShut
  {NoStop}%
\bibitem [{\citenamefont {Breoni}\ \emph {et~al.}(2023)\citenamefont {Breoni},
  \citenamefont {Kurzthaler}, \citenamefont {Liebchen}, \citenamefont
  {Löwen},\ and\ \citenamefont {Mandal}}]{Davide2023}%
  \BibitemOpen
  \bibfield  {author} {\bibinfo {author} {\bibfnamefont {D.}~\bibnamefont
  {Breoni}}, \bibinfo {author} {\bibfnamefont {C.}~\bibnamefont {Kurzthaler}},
  \bibinfo {author} {\bibfnamefont {B.}~\bibnamefont {Liebchen}}, \bibinfo
  {author} {\bibfnamefont {H.}~\bibnamefont {Löwen}},\ and\ \bibinfo {author}
  {\bibfnamefont {S.}~\bibnamefont {Mandal}},\ }\bibfield  {title} {\bibinfo
  {title} {Giant activity-induced stress plateau in entangled polymer
  solutions},\ }\href@noop {} {\bibfield  {journal} {\bibinfo  {journal}
  {arXiv}\ ,\ \bibinfo {pages} {2310.02929}} (\bibinfo {year}
  {2023})}\BibitemShut {NoStop}%
\bibitem [{\citenamefont {Moreira}\ \emph {et~al.}(2015)\citenamefont
  {Moreira}, \citenamefont {Zhang}, \citenamefont {Müller}, \citenamefont
  {Stuehn},\ and\ \citenamefont {Kremer}}]{Moreira2015}%
  \BibitemOpen
  \bibfield  {author} {\bibinfo {author} {\bibfnamefont {L.~A.}\ \bibnamefont
  {Moreira}}, \bibinfo {author} {\bibfnamefont {G.}~\bibnamefont {Zhang}},
  \bibinfo {author} {\bibfnamefont {F.}~\bibnamefont {Müller}}, \bibinfo
  {author} {\bibfnamefont {T.}~\bibnamefont {Stuehn}},\ and\ \bibinfo {author}
  {\bibfnamefont {K.}~\bibnamefont {Kremer}},\ }\bibfield  {title} {\bibinfo
  {title} {Direct equilibration and characterization of polymer melts for
  computer simulations},\ }\href
  {https://doi.org/https://doi.org/10.1002/mats.201500013} {\bibfield
  {journal} {\bibinfo  {journal} {Macromol. Theor. Simul.}\ }\textbf {\bibinfo
  {volume} {24}},\ \bibinfo {pages} {419} (\bibinfo {year} {2015})}\BibitemShut
  {NoStop}%
\bibitem [{\citenamefont {Svaneborg}\ and\ \citenamefont
  {Everaers}(2020)}]{Svaneborg2020}%
  \BibitemOpen
  \bibfield  {author} {\bibinfo {author} {\bibfnamefont {C.}~\bibnamefont
  {Svaneborg}}\ and\ \bibinfo {author} {\bibfnamefont {R.}~\bibnamefont
  {Everaers}},\ }\bibfield  {title} {\bibinfo {title} {Characteristic time and
  length scales in melts of kremer--grest bead--spring polymers with wormlike
  bending stiffness},\ }\href@noop {} {\bibfield  {journal} {\bibinfo
  {journal} {Macromolecules}\ }\textbf {\bibinfo {volume} {53}},\ \bibinfo
  {pages} {1917} (\bibinfo {year} {2020})}\BibitemShut {NoStop}%
\bibitem [{\citenamefont {Hoy}\ and\ \citenamefont {Kr\"oger}(2020)}]{Hoy2020}%
  \BibitemOpen
  \bibfield  {author} {\bibinfo {author} {\bibfnamefont {R.~S.}\ \bibnamefont
  {Hoy}}\ and\ \bibinfo {author} {\bibfnamefont {M.}~\bibnamefont {Kr\"oger}},\
  }\bibfield  {title} {\bibinfo {title} {Unified analytic expressions for the
  entanglement length, tube diameter, and plateau modulus of polymer melts},\
  }\href {https://doi.org/10.1103/PhysRevLett.124.147801} {\bibfield  {journal}
  {\bibinfo  {journal} {Phys. Rev. Lett.}\ }\textbf {\bibinfo {volume} {124}},\
  \bibinfo {pages} {147801} (\bibinfo {year} {2020})}\BibitemShut {NoStop}%
\bibitem [{\citenamefont {Milner}(2020)}]{Milner2020}%
  \BibitemOpen
  \bibfield  {author} {\bibinfo {author} {\bibfnamefont {S.~T.}\ \bibnamefont
  {Milner}},\ }\bibfield  {title} {\bibinfo {title} {Unified entanglement
  scaling for flexible, semiflexible, and stiff polymer melts and solutions},\
  }\href {https://doi.org/10.1021/acs.macromol.9b02684} {\bibfield  {journal}
  {\bibinfo  {journal} {Macromolecules}\ }\textbf {\bibinfo {volume} {53}},\
  \bibinfo {pages} {1314} (\bibinfo {year} {2020})}\BibitemShut {NoStop}%
\bibitem [{\citenamefont {Uchida}\ \emph {et~al.}(2008)\citenamefont {Uchida},
  \citenamefont {Grest},\ and\ \citenamefont {Everaers}}]{Everaers2008}%
  \BibitemOpen
  \bibfield  {author} {\bibinfo {author} {\bibfnamefont {N.}~\bibnamefont
  {Uchida}}, \bibinfo {author} {\bibfnamefont {G.~S.}\ \bibnamefont {Grest}},\
  and\ \bibinfo {author} {\bibfnamefont {R.}~\bibnamefont {Everaers}},\
  }\bibfield  {title} {\bibinfo {title} {{Viscoelasticity and primitive path
  analysis of entangled polymer liquids: From F-actin to polyethylene}},\
  }\href {https://doi.org/10.1063/1.2825597} {\bibfield  {journal} {\bibinfo
  {journal} {J. Chem. Phys.}\ }\textbf {\bibinfo {volume} {128}},\ \bibinfo
  {pages} {044902} (\bibinfo {year} {2008})}\BibitemShut {NoStop}%
\bibitem [{\citenamefont {Gutierrez-Escribano}\ \emph
  {et~al.}(2020)\citenamefont {Gutierrez-Escribano}, \citenamefont {Hormeño},
  \citenamefont {Madariaga-Marcos}, \citenamefont {Solé-Soler}, \citenamefont
  {O’Reilly}, \citenamefont {Morris}, \citenamefont {Aicart-Ramos},
  \citenamefont {Aramayo}, \citenamefont {Montoya}, \citenamefont {Kramer},
  \citenamefont {Rappsilber}, \citenamefont {Torres-Rosell}, \citenamefont
  {Moreno-Herrero},\ and\ \citenamefont {Aragon}}]{Pilar2020}%
  \BibitemOpen
  \bibfield  {author} {\bibinfo {author} {\bibfnamefont {P.}~\bibnamefont
  {Gutierrez-Escribano}}, \bibinfo {author} {\bibfnamefont {S.}~\bibnamefont
  {Hormeño}}, \bibinfo {author} {\bibfnamefont {J.}~\bibnamefont
  {Madariaga-Marcos}}, \bibinfo {author} {\bibfnamefont {R.}~\bibnamefont
  {Solé-Soler}}, \bibinfo {author} {\bibfnamefont {F.~J.}\ \bibnamefont
  {O’Reilly}}, \bibinfo {author} {\bibfnamefont {K.}~\bibnamefont {Morris}},
  \bibinfo {author} {\bibfnamefont {C.}~\bibnamefont {Aicart-Ramos}}, \bibinfo
  {author} {\bibfnamefont {R.}~\bibnamefont {Aramayo}}, \bibinfo {author}
  {\bibfnamefont {A.}~\bibnamefont {Montoya}}, \bibinfo {author} {\bibfnamefont
  {H.}~\bibnamefont {Kramer}}, \bibinfo {author} {\bibfnamefont
  {J.}~\bibnamefont {Rappsilber}}, \bibinfo {author} {\bibfnamefont
  {J.}~\bibnamefont {Torres-Rosell}}, \bibinfo {author} {\bibfnamefont
  {F.}~\bibnamefont {Moreno-Herrero}},\ and\ \bibinfo {author} {\bibfnamefont
  {L.}~\bibnamefont {Aragon}},\ }\bibfield  {title} {\bibinfo {title} {Purified
  smc5/6 complex exhibits dna substrate recognition and compaction},\ }\href
  {https://doi.org/https://doi.org/10.1016/j.molcel.2020.11.012} {\bibfield
  {journal} {\bibinfo  {journal} {Molecular Cell}\ }\textbf {\bibinfo {volume}
  {80}},\ \bibinfo {pages} {1039} (\bibinfo {year} {2020})}\BibitemShut
  {NoStop}%
\bibitem [{\citenamefont {Clapier}\ \emph {et~al.}(2017)\citenamefont
  {Clapier}, \citenamefont {Iwasa}, \citenamefont {Cairns},\ and\ \citenamefont
  {Peterson}}]{Clapier2017}%
  \BibitemOpen
  \bibfield  {author} {\bibinfo {author} {\bibfnamefont {C.~R.}\ \bibnamefont
  {Clapier}}, \bibinfo {author} {\bibfnamefont {J.}~\bibnamefont {Iwasa}},
  \bibinfo {author} {\bibfnamefont {B.~R.}\ \bibnamefont {Cairns}},\ and\
  \bibinfo {author} {\bibfnamefont {C.~L.}\ \bibnamefont {Peterson}},\
  }\bibfield  {title} {\bibinfo {title} {Mechanisms of action and regulation of
  atp-dependent chromatin-remodelling complexes},\ }\href
  {https://doi.org/10.1038/nrm.2017.26} {\bibfield  {journal} {\bibinfo
  {journal} {Nat. Rev. Mol. Cell Bio.}\ }\textbf {\bibinfo {volume} {18}},\
  \bibinfo {pages} {407} (\bibinfo {year} {2017})}\BibitemShut {NoStop}%
\bibitem [{\citenamefont {Takaki}\ \emph {et~al.}(2021)\citenamefont {Takaki},
  \citenamefont {Dey}, \citenamefont {Shi},\ and\ \citenamefont
  {Thirumalai}}]{Thirumalai2021}%
  \BibitemOpen
  \bibfield  {author} {\bibinfo {author} {\bibfnamefont {R.}~\bibnamefont
  {Takaki}}, \bibinfo {author} {\bibfnamefont {A.}~\bibnamefont {Dey}},
  \bibinfo {author} {\bibfnamefont {G.}~\bibnamefont {Shi}},\ and\ \bibinfo
  {author} {\bibfnamefont {D.}~\bibnamefont {Thirumalai}},\ }\bibfield  {title}
  {\bibinfo {title} {Theory and simulations of condensin mediated loop
  extrusion in dna},\ }\href {https://doi.org/10.1038/s41467-021-26167-1}
  {\bibfield  {journal} {\bibinfo  {journal} {Nat. Comm.}\ }\textbf {\bibinfo
  {volume} {12}},\ \bibinfo {pages} {5865} (\bibinfo {year}
  {2021})}\BibitemShut {NoStop}%
\bibitem [{\citenamefont {Kremer}\ and\ \citenamefont
  {Grest}(1990)}]{Kremer1990}%
  \BibitemOpen
  \bibfield  {author} {\bibinfo {author} {\bibfnamefont {K.}~\bibnamefont
  {Kremer}}\ and\ \bibinfo {author} {\bibfnamefont {G.~S.}\ \bibnamefont
  {Grest}},\ }\bibfield  {title} {\bibinfo {title} {{Dynamics of entangled
  linear polymer melts: A molecular-dynamics simulation}},\ }\href
  {https://doi.org/10.1063/1.458541} {\bibfield  {journal} {\bibinfo  {journal}
  {J. Chem. Phys.}\ }\textbf {\bibinfo {volume} {92}},\ \bibinfo {pages} {5057}
  (\bibinfo {year} {1990})}\BibitemShut {NoStop}%
\bibitem [{\citenamefont {Fetters}\ \emph {et~al.}(1994)\citenamefont
  {Fetters}, \citenamefont {Lohse}, \citenamefont {Richter}, \citenamefont
  {Witten},\ and\ \citenamefont {Zirkel}}]{Fetters1994}%
  \BibitemOpen
  \bibfield  {author} {\bibinfo {author} {\bibfnamefont {L.~J.}\ \bibnamefont
  {Fetters}}, \bibinfo {author} {\bibfnamefont {D.~J.}\ \bibnamefont {Lohse}},
  \bibinfo {author} {\bibfnamefont {D.}~\bibnamefont {Richter}}, \bibinfo
  {author} {\bibfnamefont {T.~A.}\ \bibnamefont {Witten}},\ and\ \bibinfo
  {author} {\bibfnamefont {A.}~\bibnamefont {Zirkel}},\ }\bibfield  {title}
  {\bibinfo {title} {Connection between polymer molecular weight, density,
  chain dimensions, and melt viscoelastic properties},\ }\href
  {https://doi.org/10.1021/ma00095a001} {\bibfield  {journal} {\bibinfo
  {journal} {Macromolecules}\ }\textbf {\bibinfo {volume} {27}},\ \bibinfo
  {pages} {4639} (\bibinfo {year} {1994})}\BibitemShut {NoStop}%
\bibitem [{\citenamefont {Kr\"oger}\ and\ \citenamefont
  {Hess}(2000)}]{Martin2000}%
  \BibitemOpen
  \bibfield  {author} {\bibinfo {author} {\bibfnamefont {M.}~\bibnamefont
  {Kr\"oger}}\ and\ \bibinfo {author} {\bibfnamefont {S.}~\bibnamefont
  {Hess}},\ }\bibfield  {title} {\bibinfo {title} {Rheological evidence for a
  dynamical crossover in polymer melts via nonequilibrium molecular dynamics},\
  }\href {https://doi.org/10.1103/PhysRevLett.85.1128} {\bibfield  {journal}
  {\bibinfo  {journal} {Phys. Rev. Lett.}\ }\textbf {\bibinfo {volume} {85}},\
  \bibinfo {pages} {1128} (\bibinfo {year} {2000})}\BibitemShut {NoStop}%
\bibitem [{\citenamefont {Fetters}\ \emph {et~al.}(1999)\citenamefont
  {Fetters}, \citenamefont {Lohse}, \citenamefont {Milner},\ and\ \citenamefont
  {Graessley}}]{Milner1999}%
  \BibitemOpen
  \bibfield  {author} {\bibinfo {author} {\bibfnamefont {L.~J.}\ \bibnamefont
  {Fetters}}, \bibinfo {author} {\bibfnamefont {D.~J.}\ \bibnamefont {Lohse}},
  \bibinfo {author} {\bibfnamefont {S.~T.}\ \bibnamefont {Milner}},\ and\
  \bibinfo {author} {\bibfnamefont {W.~W.}\ \bibnamefont {Graessley}},\
  }\bibfield  {title} {\bibinfo {title} {Packing length influence in linear
  polymer melts on the entanglement, critical, and reptation molecular
  weights},\ }\href {https://doi.org/10.1021/ma990620o} {\bibfield  {journal}
  {\bibinfo  {journal} {Macromolecules}\ }\textbf {\bibinfo {volume} {32}},\
  \bibinfo {pages} {6847} (\bibinfo {year} {1999})}\BibitemShut {NoStop}%
\bibitem [{\citenamefont {Deblais}\ \emph {et~al.}(2020)\citenamefont
  {Deblais}, \citenamefont {Maggs}, \citenamefont {Bonn},\ and\ \citenamefont
  {Woutersen}}]{Deblais2020}%
  \BibitemOpen
  \bibfield  {author} {\bibinfo {author} {\bibfnamefont {A.}~\bibnamefont
  {Deblais}}, \bibinfo {author} {\bibfnamefont {A.~C.}\ \bibnamefont {Maggs}},
  \bibinfo {author} {\bibfnamefont {D.}~\bibnamefont {Bonn}},\ and\ \bibinfo
  {author} {\bibfnamefont {S.}~\bibnamefont {Woutersen}},\ }\bibfield  {title}
  {\bibinfo {title} {Phase separation by entanglement of active polymerlike
  worms},\ }\href {https://doi.org/10.1103/PhysRevLett.124.208006} {\bibfield
  {journal} {\bibinfo  {journal} {Phys. Rev. Lett.}\ }\textbf {\bibinfo
  {volume} {124}},\ \bibinfo {pages} {208006} (\bibinfo {year}
  {2020})}\BibitemShut {NoStop}%
\bibitem [{\citenamefont {Plimpton}(1995)}]{Plimpton1995}%
  \BibitemOpen
  \bibfield  {author} {\bibinfo {author} {\bibfnamefont {S.}~\bibnamefont
  {Plimpton}},\ }\bibfield  {title} {\bibinfo {title} {Fast parallel algorithms
  for short-range molecular dynamics},\ }\href
  {https://doi.org/https://doi.org/10.1006/jcph.1995.1039} {\bibfield
  {journal} {\bibinfo  {journal} {J. Comp. Phys.}\ }\textbf {\bibinfo {volume}
  {117}},\ \bibinfo {pages} {1} (\bibinfo {year} {1995})}\BibitemShut {NoStop}%
\bibitem [{\citenamefont {Ilker}\ \emph {et~al.}(2021)\citenamefont {Ilker},
  \citenamefont {Castellana},\ and\ \citenamefont {Joanny}}]{Ilker2021}%
  \BibitemOpen
  \bibfield  {author} {\bibinfo {author} {\bibfnamefont {E.}~\bibnamefont
  {Ilker}}, \bibinfo {author} {\bibfnamefont {M.}~\bibnamefont {Castellana}},\
  and\ \bibinfo {author} {\bibfnamefont {J.-F. m.~c.}\ \bibnamefont {Joanny}},\
  }\bibfield  {title} {\bibinfo {title} {Long-time diffusion and energy
  transfer in polydisperse mixtures of particles with different temperatures},\
  }\href {https://doi.org/10.1103/PhysRevResearch.3.023207} {\bibfield
  {journal} {\bibinfo  {journal} {Phys. Rev. Res.}\ }\textbf {\bibinfo {volume}
  {3}},\ \bibinfo {pages} {023207} (\bibinfo {year} {2021})}\BibitemShut
  {NoStop}%
\bibitem [{\citenamefont {Doi}\ and\ \citenamefont {Edwards}(1986)}]{Doi1986}%
  \BibitemOpen
  \bibfield  {author} {\bibinfo {author} {\bibfnamefont {M.}~\bibnamefont
  {Doi}}\ and\ \bibinfo {author} {\bibfnamefont {S.~F.}\ \bibnamefont
  {Edwards}},\ }\href@noop {} {\emph {\bibinfo {title} {The Theory of Polymer
  Dynamics}}}\ (\bibinfo  {publisher} {Oxford University Press},\ \bibinfo
  {year} {1986})\BibitemShut {NoStop}%
\bibitem [{\citenamefont {Wang}(2003)}]{Wang2003}%
  \BibitemOpen
  \bibfield  {author} {\bibinfo {author} {\bibfnamefont {S.-Q.}\ \bibnamefont
  {Wang}},\ }\bibfield  {title} {\bibinfo {title} {Chain dynamics in entangled
  polymers: Diffusion versus rheology and their comparison},\ }\href@noop {}
  {\bibfield  {journal} {\bibinfo  {journal} {J. Polym. Sci. Pol. Phys.}\
  }\textbf {\bibinfo {volume} {41}},\ \bibinfo {pages} {1589} (\bibinfo {year}
  {2003})}\BibitemShut {NoStop}%
\bibitem [{\citenamefont {Kr{\"o}ger}\ \emph {et~al.}(2023)\citenamefont
  {Kr{\"o}ger}, \citenamefont {Dietz}, \citenamefont {Hoy},\ and\ \citenamefont
  {Luap}}]{Kroger2023}%
  \BibitemOpen
  \bibfield  {author} {\bibinfo {author} {\bibfnamefont {M.}~\bibnamefont
  {Kr{\"o}ger}}, \bibinfo {author} {\bibfnamefont {J.~D.}\ \bibnamefont
  {Dietz}}, \bibinfo {author} {\bibfnamefont {R.~S.}\ \bibnamefont {Hoy}},\
  and\ \bibinfo {author} {\bibfnamefont {C.}~\bibnamefont {Luap}},\ }\bibfield
  {title} {\bibinfo {title} {The z1+ package: Shortest multiple disconnected
  path for the analysis of entanglements in macromolecular systems},\
  }\href@noop {} {\bibfield  {journal} {\bibinfo  {journal} {Comput. Phys.
  Commun.}\ }\textbf {\bibinfo {volume} {283}},\ \bibinfo {pages} {108567}
  (\bibinfo {year} {2023})}\BibitemShut {NoStop}%
\bibitem [{\citenamefont {Halverson}\ \emph {et~al.}(2011)\citenamefont
  {Halverson}, \citenamefont {Lee}, \citenamefont {Grest}, \citenamefont
  {Grosberg},\ and\ \citenamefont {Kremer}}]{Halverson2011}%
  \BibitemOpen
  \bibfield  {author} {\bibinfo {author} {\bibfnamefont {J.~D.}\ \bibnamefont
  {Halverson}}, \bibinfo {author} {\bibfnamefont {W.~B.}\ \bibnamefont {Lee}},
  \bibinfo {author} {\bibfnamefont {G.~S.}\ \bibnamefont {Grest}}, \bibinfo
  {author} {\bibfnamefont {A.~Y.}\ \bibnamefont {Grosberg}},\ and\ \bibinfo
  {author} {\bibfnamefont {K.}~\bibnamefont {Kremer}},\ }\bibfield  {title}
  {\bibinfo {title} {{Molecular dynamics simulation study of nonconcatenated
  ring polymers in a melt. II. Dynamics}},\ }\href
  {https://doi.org/10.1063/1.3587138} {\bibfield  {journal} {\bibinfo
  {journal} {J. Chem. Phys.}\ }\textbf {\bibinfo {volume} {134}},\ \bibinfo
  {pages} {204905} (\bibinfo {year} {2011})}\BibitemShut {NoStop}%
\bibitem [{\citenamefont {Herrmann}\ \emph {et~al.}(2012)\citenamefont
  {Herrmann}, \citenamefont {Kresse}, \citenamefont {Wohlfahrt}, \citenamefont
  {Bauer}, \citenamefont {Privalov}, \citenamefont {Kruk}, \citenamefont
  {Fatkullin}, \citenamefont {Fujara},\ and\ \citenamefont
  {Rössler}}]{Herrmann2012}%
  \BibitemOpen
  \bibfield  {author} {\bibinfo {author} {\bibfnamefont {A.}~\bibnamefont
  {Herrmann}}, \bibinfo {author} {\bibfnamefont {B.}~\bibnamefont {Kresse}},
  \bibinfo {author} {\bibfnamefont {M.}~\bibnamefont {Wohlfahrt}}, \bibinfo
  {author} {\bibfnamefont {I.}~\bibnamefont {Bauer}}, \bibinfo {author}
  {\bibfnamefont {A.~F.}\ \bibnamefont {Privalov}}, \bibinfo {author}
  {\bibfnamefont {D.}~\bibnamefont {Kruk}}, \bibinfo {author} {\bibfnamefont
  {N.}~\bibnamefont {Fatkullin}}, \bibinfo {author} {\bibfnamefont
  {F.}~\bibnamefont {Fujara}},\ and\ \bibinfo {author} {\bibfnamefont {E.~A.}\
  \bibnamefont {Rössler}},\ }\bibfield  {title} {\bibinfo {title} {Mean square
  displacement and reorientational correlation function in entangled polymer
  melts revealed by field cycling 1h and 2h nmr relaxometry},\ }\href
  {https://doi.org/10.1021/ma301099h} {\bibfield  {journal} {\bibinfo
  {journal} {Macromolecules}\ }\textbf {\bibinfo {volume} {45}},\ \bibinfo
  {pages} {6516} (\bibinfo {year} {2012})}\BibitemShut {NoStop}%
\bibitem [{\citenamefont {Ziebert}\ and\ \citenamefont
  {Aranson}(2008)}]{Aranson2008}%
  \BibitemOpen
  \bibfield  {author} {\bibinfo {author} {\bibfnamefont {F.}~\bibnamefont
  {Ziebert}}\ and\ \bibinfo {author} {\bibfnamefont {I.~S.}\ \bibnamefont
  {Aranson}},\ }\bibfield  {title} {\bibinfo {title} {Rheological and
  structural properties of dilute active filament solutions},\ }\href
  {https://doi.org/10.1103/PhysRevE.77.011918} {\bibfield  {journal} {\bibinfo
  {journal} {Phys. Rev. E}\ }\textbf {\bibinfo {volume} {77}},\ \bibinfo
  {pages} {011918} (\bibinfo {year} {2008})}\BibitemShut {NoStop}%
\bibitem [{\citenamefont {Humphrey}\ \emph {et~al.}(2002)\citenamefont
  {Humphrey}, \citenamefont {Duggan}, \citenamefont {Saha}, \citenamefont
  {Smith},\ and\ \citenamefont {Käs}}]{Humphrey2002}%
  \BibitemOpen
  \bibfield  {author} {\bibinfo {author} {\bibfnamefont {D.}~\bibnamefont
  {Humphrey}}, \bibinfo {author} {\bibfnamefont {C.}~\bibnamefont {Duggan}},
  \bibinfo {author} {\bibfnamefont {D.}~\bibnamefont {Saha}}, \bibinfo {author}
  {\bibfnamefont {D.}~\bibnamefont {Smith}},\ and\ \bibinfo {author}
  {\bibfnamefont {J.}~\bibnamefont {Käs}},\ }\bibfield  {title} {\bibinfo
  {title} {Active fluidization of polymer networks through molecular motors},\
  }\href {https://doi.org/10.1038/416413a} {\bibfield  {journal} {\bibinfo
  {journal} {Nature}\ }\textbf {\bibinfo {volume} {416}},\ \bibinfo {pages}
  {413} (\bibinfo {year} {2002})}\BibitemShut {NoStop}%
\bibitem [{\citenamefont {Winkler}\ \emph {et~al.}(2017)\citenamefont
  {Winkler}, \citenamefont {Elgeti},\ and\ \citenamefont
  {Gompper}}]{Roland2017}%
  \BibitemOpen
  \bibfield  {author} {\bibinfo {author} {\bibfnamefont {R.~G.}\ \bibnamefont
  {Winkler}}, \bibinfo {author} {\bibfnamefont {J.}~\bibnamefont {Elgeti}},\
  and\ \bibinfo {author} {\bibfnamefont {G.}~\bibnamefont {Gompper}},\
  }\bibfield  {title} {\bibinfo {title} {Active polymers — emergent
  conformational and dynamical properties: A brief review},\ }\href
  {https://doi.org/10.7566/JPSJ.86.101014} {\bibfield  {journal} {\bibinfo
  {journal} {J. Phys. Soc. Jpn.}\ }\textbf {\bibinfo {volume} {86}},\ \bibinfo
  {pages} {101014} (\bibinfo {year} {2017})}\BibitemShut {NoStop}%
\bibitem [{\citenamefont {Anand}\ and\ \citenamefont
  {Singh}(2020)}]{Anand2020}%
  \BibitemOpen
  \bibfield  {author} {\bibinfo {author} {\bibfnamefont {S.~K.}\ \bibnamefont
  {Anand}}\ and\ \bibinfo {author} {\bibfnamefont {S.~P.}\ \bibnamefont
  {Singh}},\ }\bibfield  {title} {\bibinfo {title} {Conformation and dynamics
  of a self-avoiding active flexible polymer},\ }\href
  {https://doi.org/10.1103/PhysRevE.101.030501} {\bibfield  {journal} {\bibinfo
   {journal} {Phys. Rev. E}\ }\textbf {\bibinfo {volume} {101}},\ \bibinfo
  {pages} {030501} (\bibinfo {year} {2020})}\BibitemShut {NoStop}%
\end{thebibliography}%

\end{document}